\documentclass[aps,twocolumn,showpacs,amsmath,amssymb,prl,superscriptaddress,floatfix,longbibliography]{revtex4-2}

\usepackage{hyperref}
\usepackage{bm}
\usepackage{physics}
\usepackage{braket}
\usepackage{amsmath}
\usepackage{amssymb}
\usepackage{mathtools}
\usepackage{graphicx}
\usepackage{dcolumn}
\usepackage{bm}
\usepackage{multirow}
\usepackage{comment}
\usepackage{leftidx}
\usepackage{subcaption} 

\usepackage{color}
\usepackage{float}
\usepackage{capt-of}
\usepackage{mwe}
\usepackage{amsfonts}
\usepackage[capitalise]{cleveref}
\usepackage[shortlabels]{enumitem}
\usepackage[german,english]{babel}
\usepackage{gensymb}
\usepackage[thinc]{esdiff}
\usepackage{url}
\graphicspath{ {./Images/Free}, {./Images/bound} }
\setlist[enumerate,1]{label={(\roman*)}}
\begin{document}



\title{Effect of Rabi dynamics in resonant x-ray scattering of intense attosecond pulses}


\author{Akilesh Venkatesh}
\email[]{avenkatesh[at]anl[dot]gov}
\affiliation{Chemical Sciences and Engineering Division, Argonne National Laboratory, Lemont, Illinois 60439, USA}

\author{Phay J. Ho}
\email[]{pho[at]anl[dot]gov}
\affiliation{Chemical Sciences and Engineering Division, Argonne National Laboratory, Lemont, Illinois 60439, USA}


\date{\today}

\begin{abstract}
We theoretically study the effects of Rabi oscillations on resonant x-ray scattering in Ne\textsuperscript{+} using intense attosecond and few-femtosecond pulses. The total photon yield strongly depends on the pulse area and has an upper bound at high intensities. Resonant scattering yields with attosecond pulses can be an order of magnitude higher than those far from resonances. Interference between resonant fluorescence and elastic scattering channels depends on the pulse area and initial state. Our results suggest that resonant x-ray scattering can be exploited for high-resolution, site-specific imaging.
\end{abstract}

\pacs{}

\maketitle


The advent of x-ray free electron lasers (XFELs)\cite{XFEL1, EuropeanXFEL1, SACLA_1, swissFEL_1, XFELscience_overview_Young, XFELdevelopmentsummary_2020} has fueled the ambition to capture structures and ultrafast dynamics of isolated, non-crystalline samples in their native environment at atomic resolution using single-shot coherent diffractive imaging (CDI). This goal of single-particle imaging is driven by the concept of ``imaging before destruction'', where each intense ultrashort x-ray pulse delivers a diffraction pattern before sample damage occurs\cite{neutze2000_imagebeforedestroy}, typically at a photon energy that supports the desired resolution and far away from resonances.  Based on the linear scattering model, the non-resonant scattered signals scale linearly with incident fluence \cite{Kirz_Jacobsen_Howells_1995}.  However, the intense x-ray pulses necessary for high-resolution imaging can cause extensive electronic and structural damage to the sample initiated through sequential absorption of multiple x-ray photons, leading to changes in both absorption and scattering responses during the pulse \cite{Young2010, Rudek-2012-NatPho, Doumy-PRL-2011, Hoener-PRL-2010,Schorb-PRL-2012, Bostedt2012, Femtosecond_reduction_formfactor2023} and a reduced resolution \cite{Yumoto2022,Ekeberg2024}.

Despite efforts to improve CDI through enhanced sample delivery \cite{Rafie-Zinedine-JSR-2024}, advanced data processing \cite{Zimmermann2023}, and structural retrieval methods \cite{Ayyer:21}, single-particle imaging remains challenging. However, the arrival of the attosecond x-ray pulses offer new strategies to mitigate radiation damage. Current XFEL facilities can generate intense and nearly Fourier transform-limited few-femtosecond and attosecond pulses in both soft and hard x-ray regimes \cite{Marinelli-APL-2017, Huang-PRL-2017, attoXLEAP2020, Malyzhenkov-2020-PRR, Trebushinin-Photonics-2023}. These pulses can largely suppress structural damage and ionization from Auger decays as their duration is shorter than both the core-hole lifetime and the onset of nuclear motion. Additionally, these pulses allow exploration of x-ray-driven Rabi oscillation as its rate can exceed the inner-shell decay rate. 

\begin{figure}
\resizebox{90mm}{!}{\includegraphics{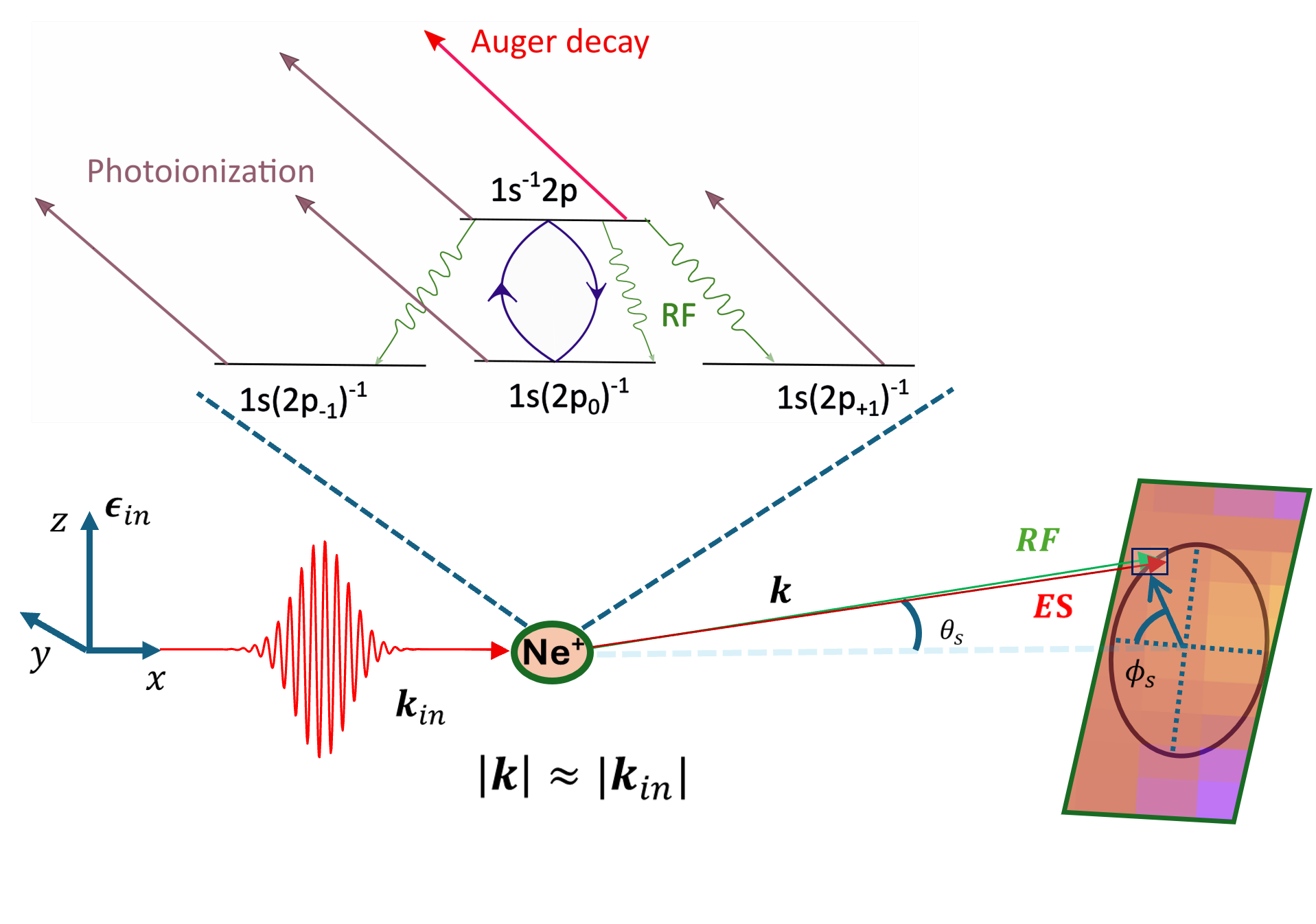}}
\caption{\label{Schematic_diagram}
Schematic diagram showing both elastic scattering (ES) and resonant fluorescence (RF) channels for the outgoing photons from a resonantly driven Ne\textsuperscript{+} atom. The inset figure shows the electronic states and two decay channels, photoionization and Auger decay.
}
\end{figure}

Recently, Rabi oscillation dynamics at short wavelengths have garnered significant attention \cite{nandi2022_observationRabidynamics, Zhang-2024-PRA,Zhang-2024-PRA2}. Direct experimental measurements in the XUV domain have been demonstrated in helium using free-electron lasers \cite{nandi2022_observationRabidynamics}. In the x-ray domain, previous theory works predicted that x-ray driven Rabi oscillations can modify both Auger electron and resonant fluorescence spectra of Ne systems.  Using a transient ("hidden") resonance in Ne\textsuperscript{+} \cite{Hiddenresonance_kanter}, the broadening of the Auger electron spectrum was measured.  An experiment reported that resonant scattering with intense femtosecond and attosecond x-ray pulses can exploit transient resonances in Xe ions to achieve enhanced scattering signals \cite{Tais_2022_Xenon_preprint} in xenon clusters. Another experiment on neon clusters using attosecond pulses \cite{attoXLEAP2020} investigated the role of Rabi oscillations in ultrafast x-ray scattering \cite{Tais_Ne_apstalk}.

In this paper, we theoretically explore the impact of x-ray driven Rabi dynamics and properties of transient resonances on the scattering response using Ne\textsuperscript{+} as a model system. As shown in Fig.~\ref{Schematic_diagram}, the scattered photon can come from either resonant fluorescence or elastic scattering channel and has energy close to incident photon energy. Our treatment goes beyond the Kramers-Heisenberg description and accounts for the interference between these two channels.  We find that Rabi oscillation can strongly modulate the scattered photon yield and
the interference is sensitive to the pulse intensity and initial state of the transient resonances.

The parameters used in this work are motivated by the range of pulse parameters used in recent experiments~\cite{Hiddenresonance_kanter, Tais_Ne_apstalk} with Ne\textsuperscript{+}. The incident x-ray photon energy  $\omega_{in}$ of 849.8 eV is chosen to be resonant with the $1s2p^{-1} \rightarrow 1s^{-1}2p$ transition.
The incident photon momentum $\boldsymbol{k}_{in}$ and the polarization $\boldsymbol{\epsilon}_{in}$ are chosen to be in $\hat{x}$ and $\hat{z}$ direction, respectively (Fig.~\ref{Schematic_diagram}). The incident pulse can drive Rabi oscillations between the core-hole excited state $1s^{-1}2p$ and the ground state $1s2p_0^{-1}$. The other degenerate ground states, $1s2p_{-1}^{-1}$ and $1s2p_{+1}^{-1}$, do not participate in the Rabi oscillations. The pulse can also ionize any of the valence electrons. The Ne\textsuperscript{+} in addition can become ionized through Auger decay from the core-hole excited state whose lifetime is about 2.4 fs~\cite{Hiddenresonance_kanter}.

Here, we explore the scattering response for pulse durations comparable or shorter than the core-hole lifetime and pulse intensities necessary to drive Rabi oscillations. The scattering response can come from both resonant fluorescence and elastic (Thomson) scattering channel.  In our QED treatment, the total vector potential is expressed as a sum of an incident classical field and an outgoing quantized field. The incident field is treated non-perturbatively~\cite{KB, NLCPRA_1} and the outgoing field is treated perturbatively to the first order~\cite{KB, NLCPRA_1, NLCPRA_4}. This is reasonable because the scattering processes of interest involve $n$ incident photons but only one outgoing photon per atom.
The ansatz for the total wavefunction is written as
\begin{equation} \label{wfn_ansatz_Nelectron}
    \ket{\psi_{total}(t) } = \psi^{(0)} (t) \ket{0} 
    + \sum_{\boldsymbol{k},\boldsymbol{\epsilon}} \psi_{\boldsymbol{k},\boldsymbol{\epsilon}} ^{(1)}(t) e^{-i \omega_{k} t} {\hat{a}_{\boldsymbol{k},\boldsymbol{\epsilon}} }^{\dagger}  \ket{0}.
\end{equation}
The first term on the right side describes the unscattered wave amplitude $\psi^{(0)} (t)$ with no outgoing photons and this accounts for the Rabi dynamics driven by a classical field.  The second term describes the scattered wave amplitude for a scattered photon with momentum $\boldsymbol{k}$ and polarization $\boldsymbol{\epsilon}$ and the entangled electronic wavefunction, $\psi_{\boldsymbol{k},\boldsymbol{\epsilon}} ^{(1)}(t)$, which enables the computation of angular distribution, photon yield and energy spectrum of the scattered photon. We restrict the discussion to the case of near-elastically scattered photons, that is, $\omega_k \approx \omega_{in}$, where $\omega_k = \abs{\boldsymbol{k}}c$. A detailed description of our approach can be found in the companion article~\cite{Companion_article}.

We first consider the Ne\textsuperscript{+} initial state to be an equal superposition of the 3 degenerate ground states ($\ket{\psi_i} = \frac{1}{\sqrt{3}} \big[ \ket{ 1s2p_{-1}^{-1} } + \ket{ 1s2p_0^{-1}} + \ket{ 1s2p_{+1}^{-1} } \big]$) shown in Fig.~\ref{Schematic_diagram}.  Then, the results for other initial states are presented. Note that previous works have experimentally demonstrated the creation and control of a coherent superposition of degenerate states in neon~\cite{Bergman_Neon_superposition_PRL2003, Bergman_Neon_superposition_PRA2007_theory, Bergman_Neon_superposition_PRA2007_exp} based on generalizations of the stimulated Raman adiabatic passage technique.

\begin{figure}[hbt!]
\vspace{0.2cm}
\begin{minipage}[t][3ex][t]{0.001\textwidth}
\vspace{-3.4cm}
\hspace{0.0cm}
(a)
\end{minipage}
\begin{minipage}[t]{0.235\textwidth}
\includegraphics[width=\linewidth]{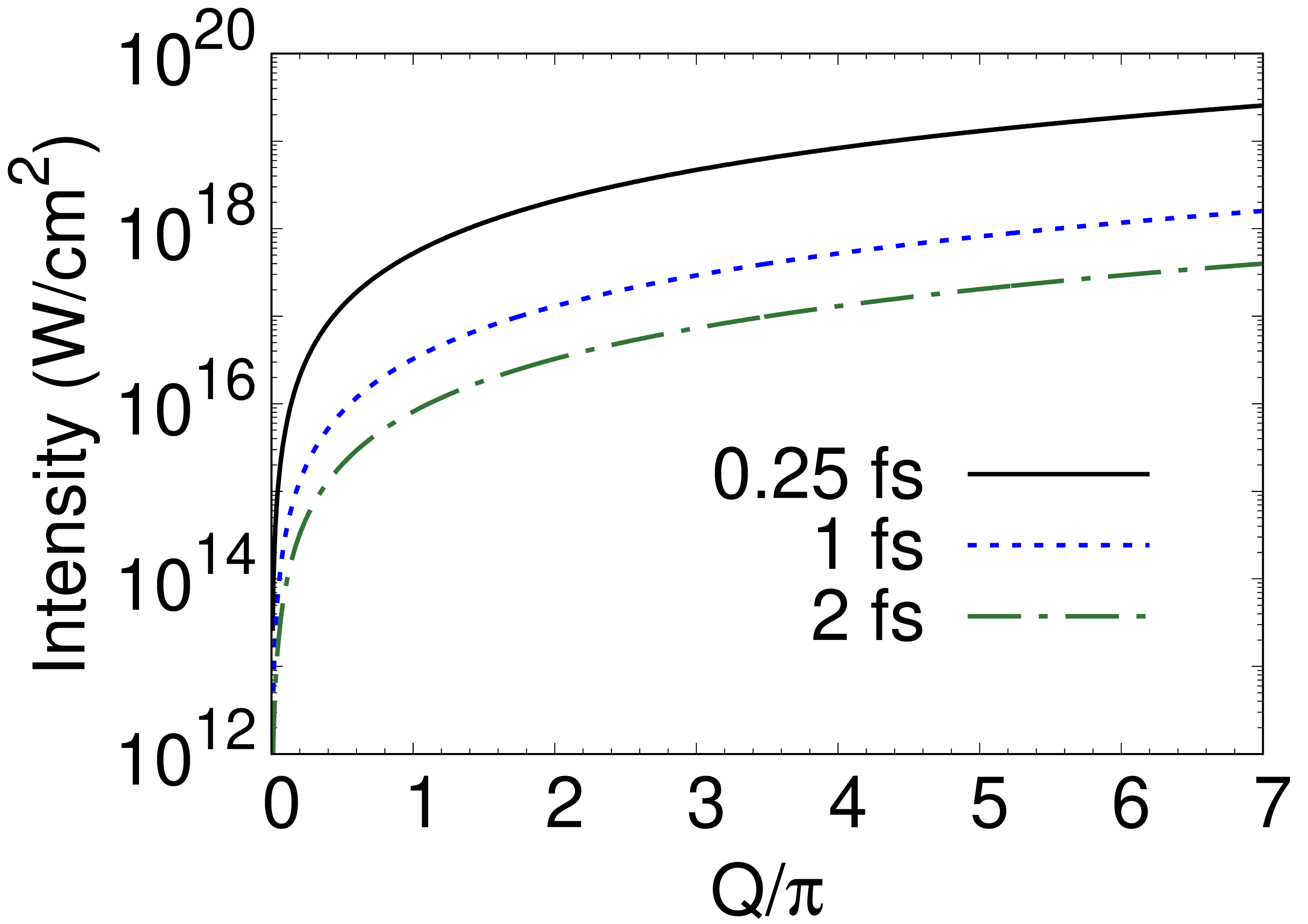}
\end{minipage}
\begin{minipage}[t][3ex][t]{0.001\textwidth}
\vspace{-3.4cm}
\hspace{-0.25cm}
(c)
\end{minipage}
\begin{minipage}[t]{0.225\textwidth}
\includegraphics[width=\linewidth]{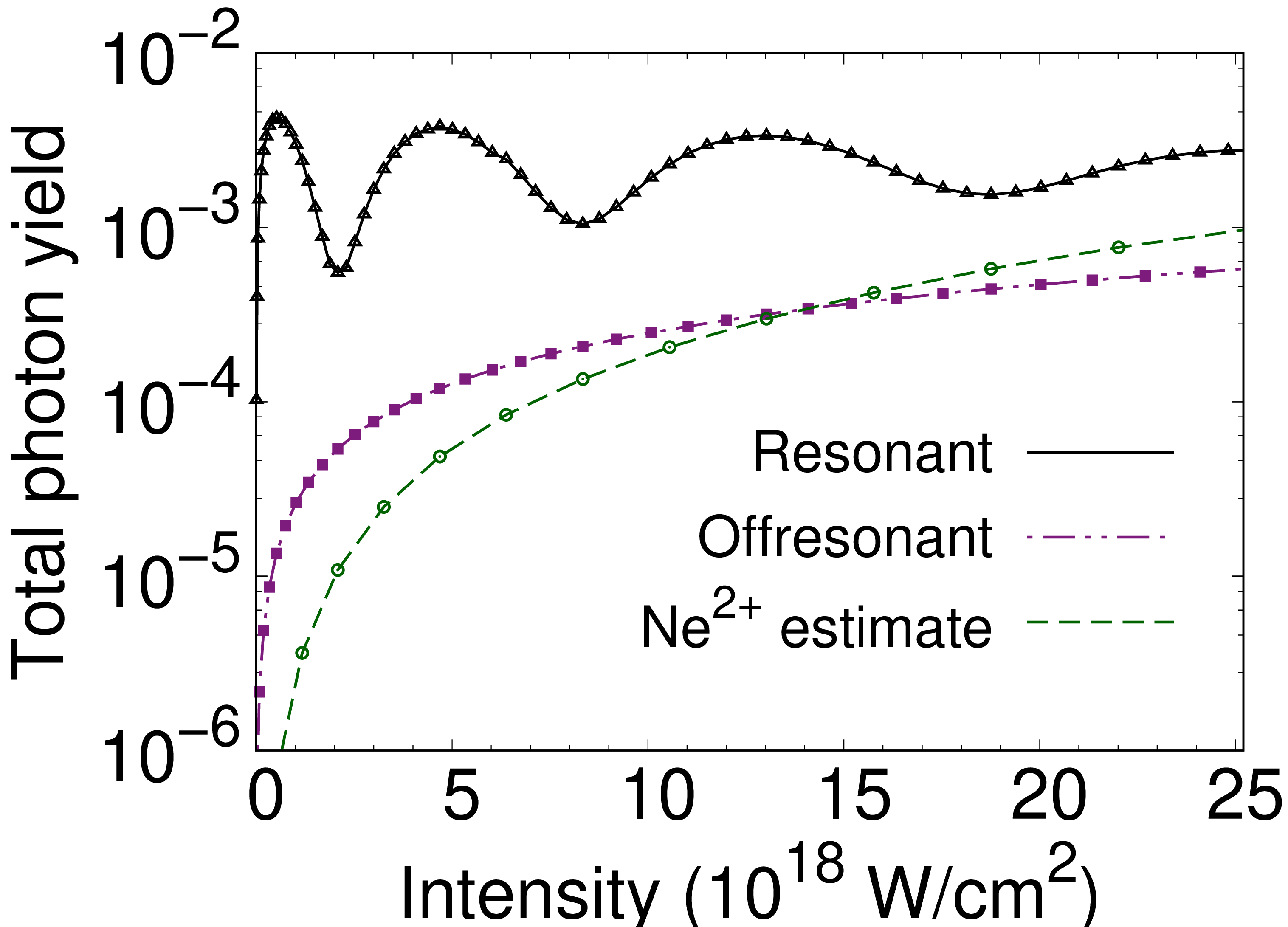}
\end{minipage}
\begin{subfigure}{0.8\linewidth}
    \vspace{-0.5cm}
    \begin{minipage}[t][3ex][t]{0.01\textwidth}
        (b) 
    \end{minipage}
    \vspace{-0.4cm}
    \begin{minipage}[b]{0.95\textwidth}
    \hspace{3 pt}
    \end{minipage}
    \begin{minipage}[t]{0.95\textwidth}
        \vspace{0cm}
        \includegraphics[width=\linewidth]{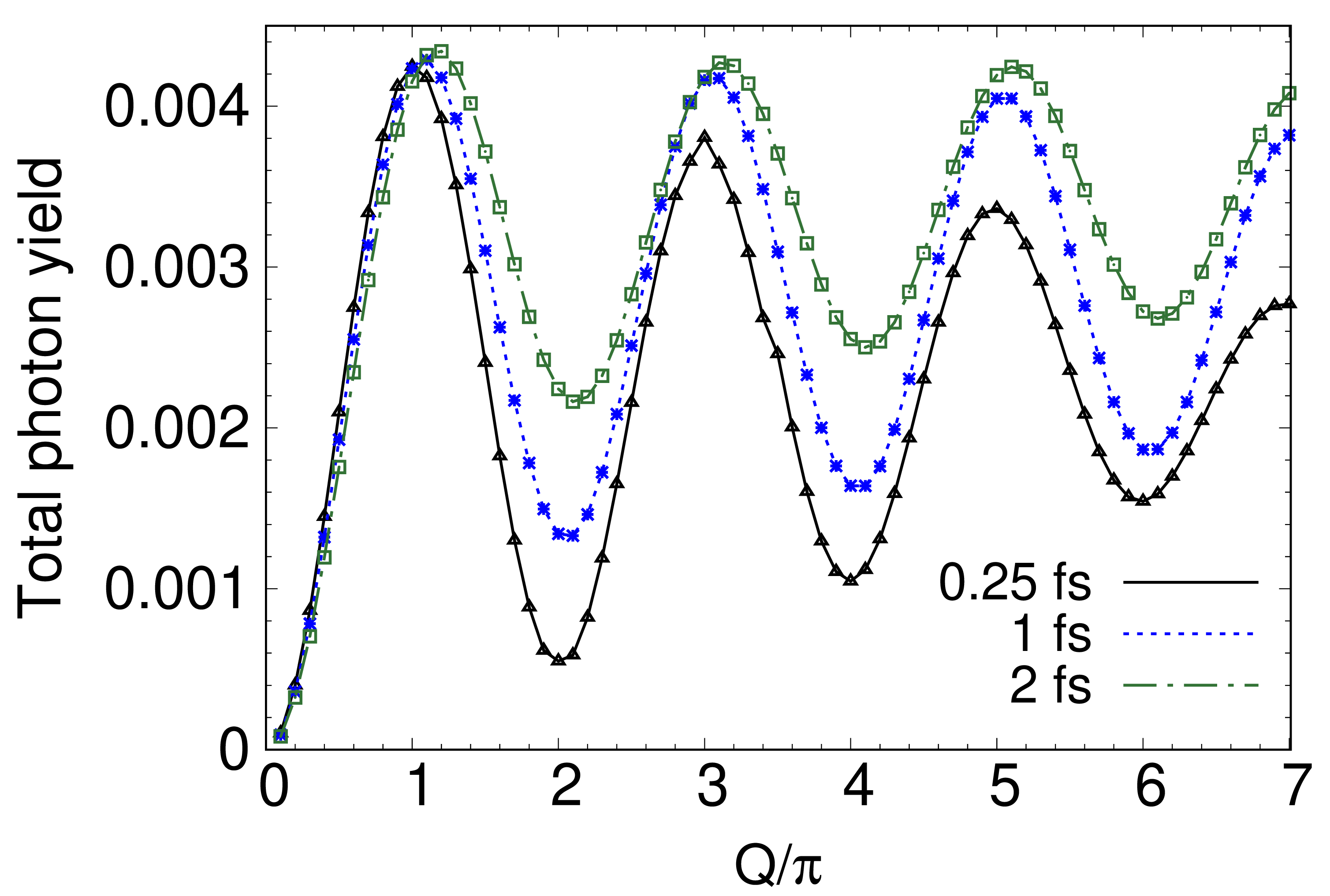}
    \end{minipage}
\end{subfigure}
\caption{(a) Pulse intensity as a function of pulse area, Q, for Ne$^+$ for different pulse durations. (b) Total photon yield per atom as a function of Q for different pulse durations. (c) Comparison of the total photon yield from resonant and non-resonant ($\omega_{in}\sim2.7$ keV) scattering from Ne\textsuperscript{+} and estimated scattering from the Ne\textsuperscript{2+} population formed during the pulse.
}
\label{Fig_signal_pulesarea}
\end{figure}

The Rabi oscillation dynamics depends on the pulse area ~\cite{Pulsearea_defn_Eberly, Cavaletto_ResFluor_PRA}, $Q = \int_{-\infty}^{\infty} \Omega(t) dt$.
Here $\Omega(t)$ is the Rabi frequency, which is given by the dot product of electric field and the dipole moment of the resonant transition.
From the area theorem~\cite{Pulsearea_defn_Eberly, pulseareatheorem_2018, Cavaletto_ResFluor_PRA}, for 2$\pi$-type pulses ($Q = 2n\pi$), there is no population in the excited state after the pulse. For $\pi$-type pulses ($Q = (2n+1) \pi$), the entire ground state ($1s2p_0^{-1}$) population appears in the excited state after the pulse. Since resonant fluorescence depends on the excited-state population, its temporal emission profile, as well as its interference with elastic scattering, can be controlled by the pulse area. Fig.~\ref{Fig_signal_pulesarea}(a) shows the range of intensities required to explore the pulse areas discussed in this work for different pulse durations with the highest intensity of $2.6 \times 10^{19}$ W/cm$^2$ to reach Q = 7$\pi$ with a 0.25 fs pulse. This intensity range is achievable with current XFELs.

\begin{figure*}
\begin{minipage}[b][][b]{0.245\textwidth}
\includegraphics[width=\textwidth]{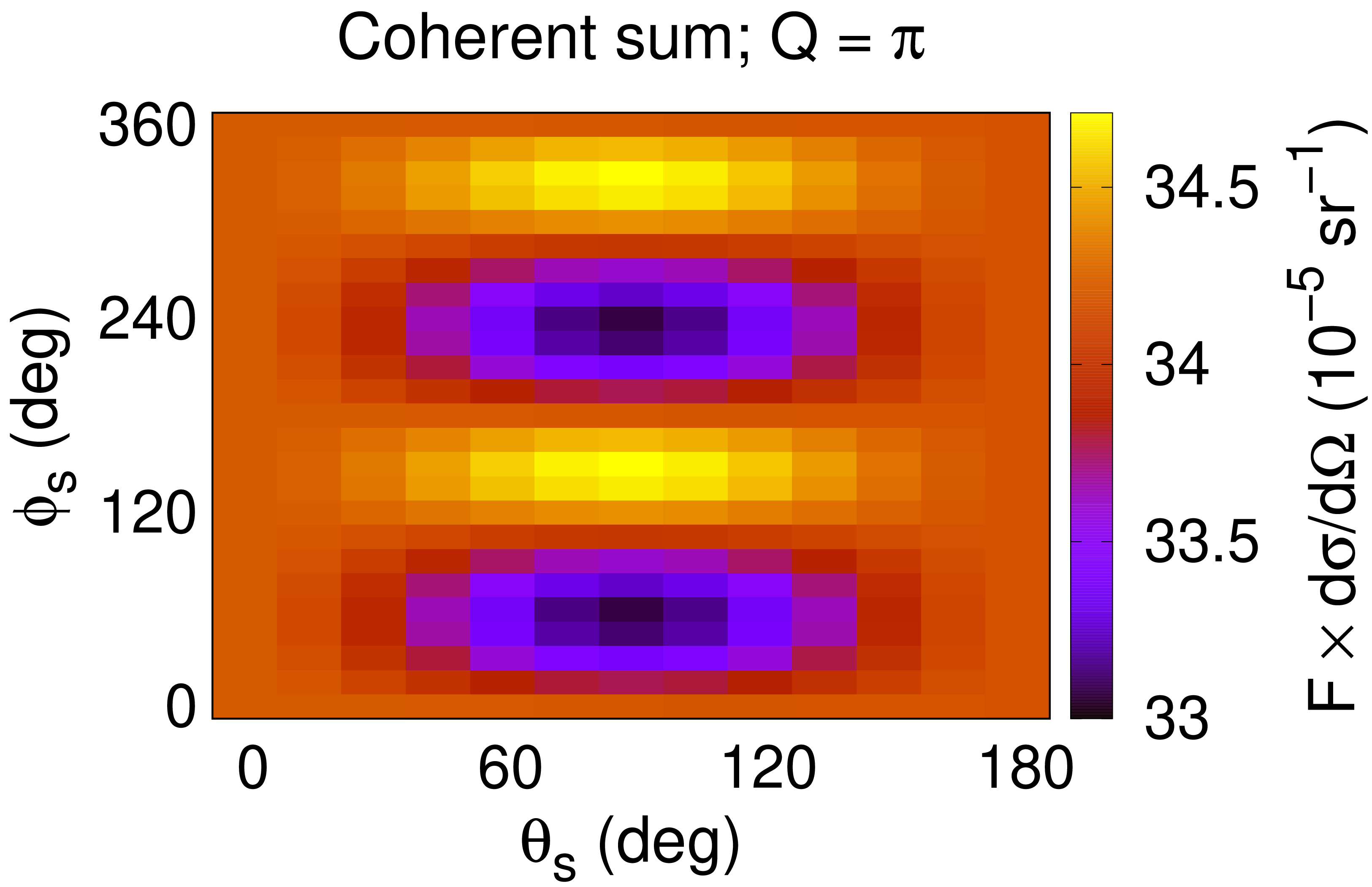}
\end{minipage}
\hspace{0.03cm} 
\begin{minipage}[b][][b]{0.235\textwidth}
\includegraphics[width=\textwidth]{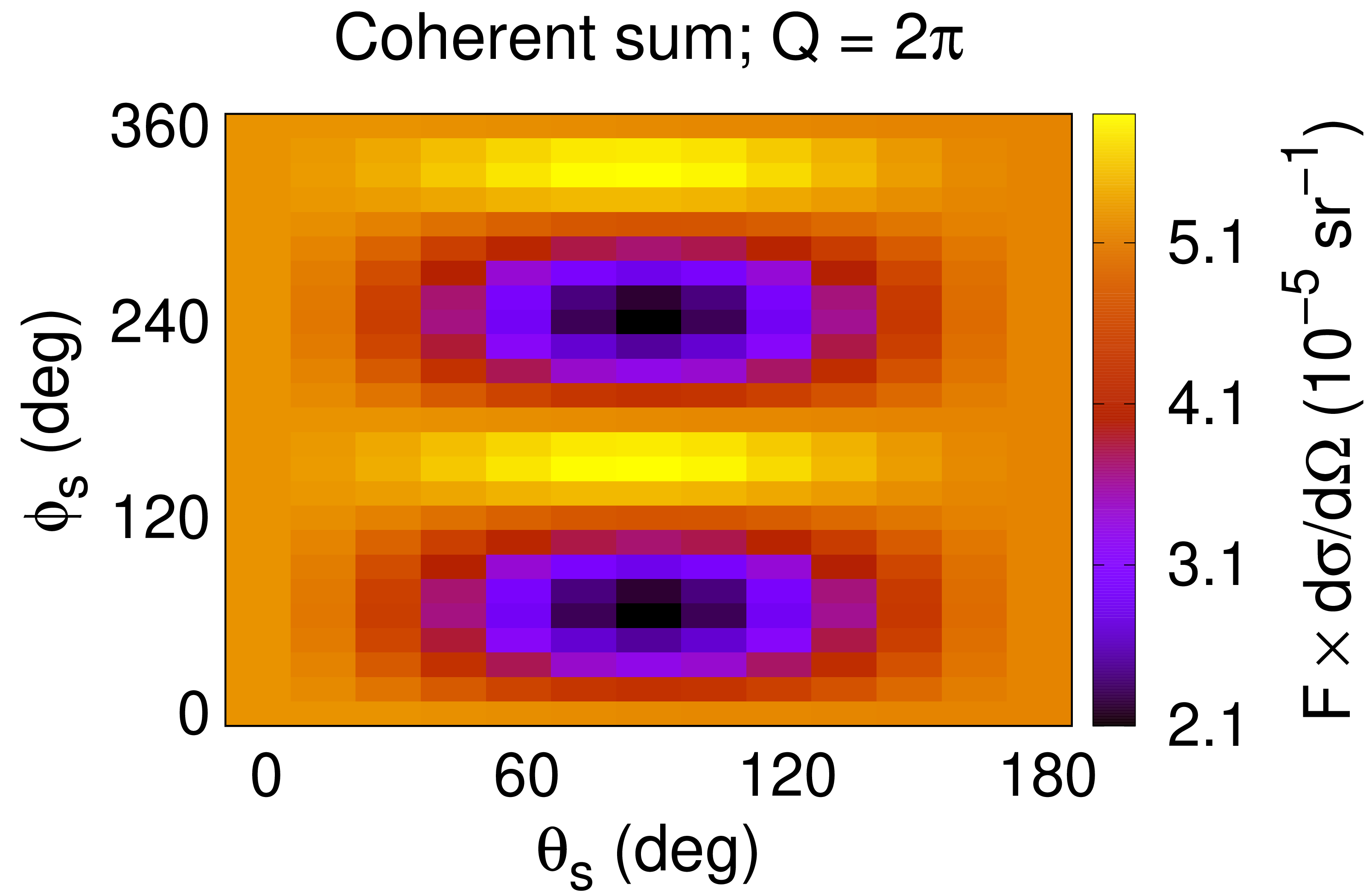}
\end{minipage}
\hspace{0.03cm}
\begin{minipage}[b][][b]{0.235\textwidth}
\includegraphics[width=\textwidth]{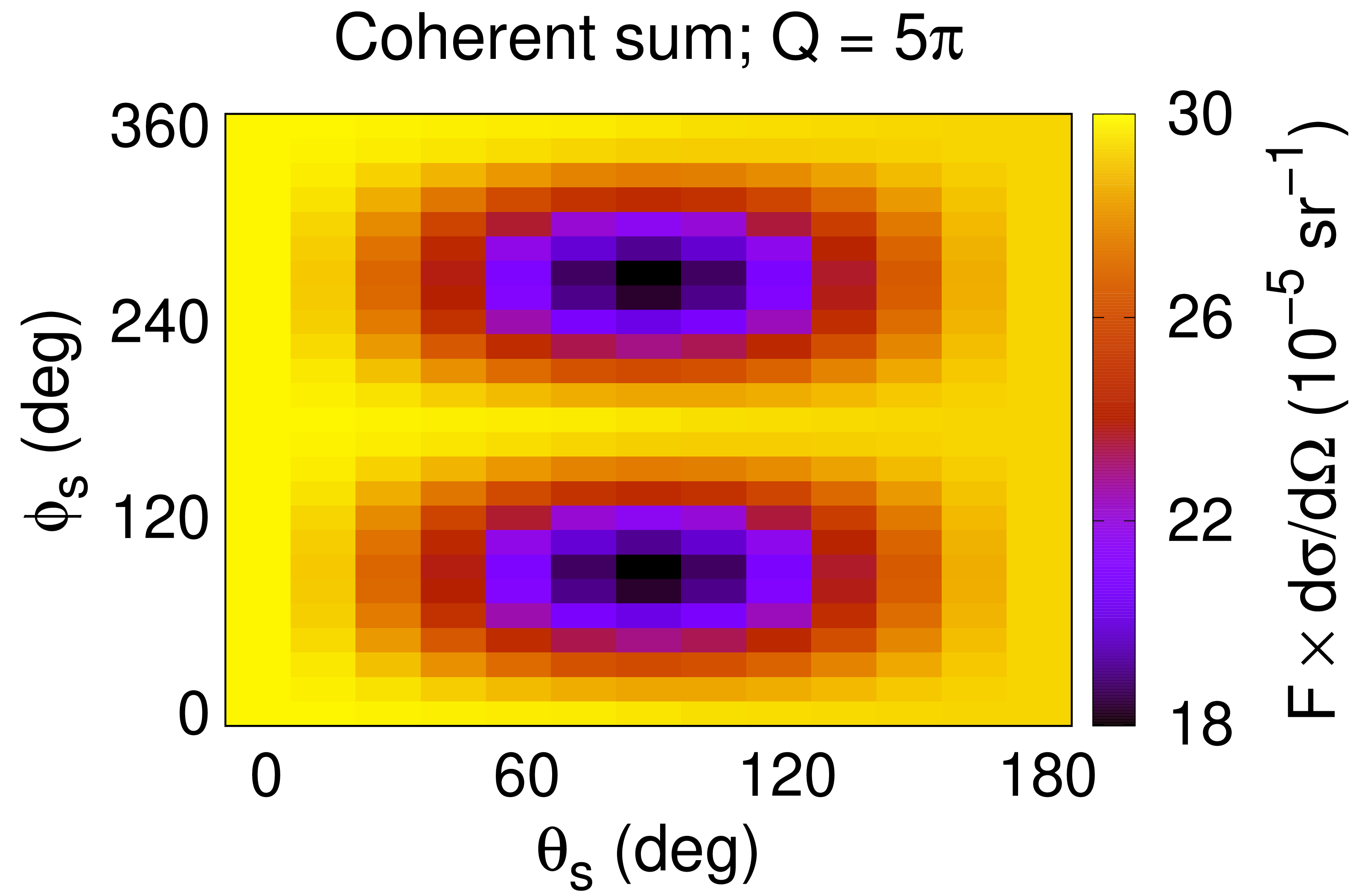}
\end{minipage}
\hspace{0.03cm}
\begin{minipage}[b][][b]{0.245\textwidth}
\includegraphics[width=\textwidth]{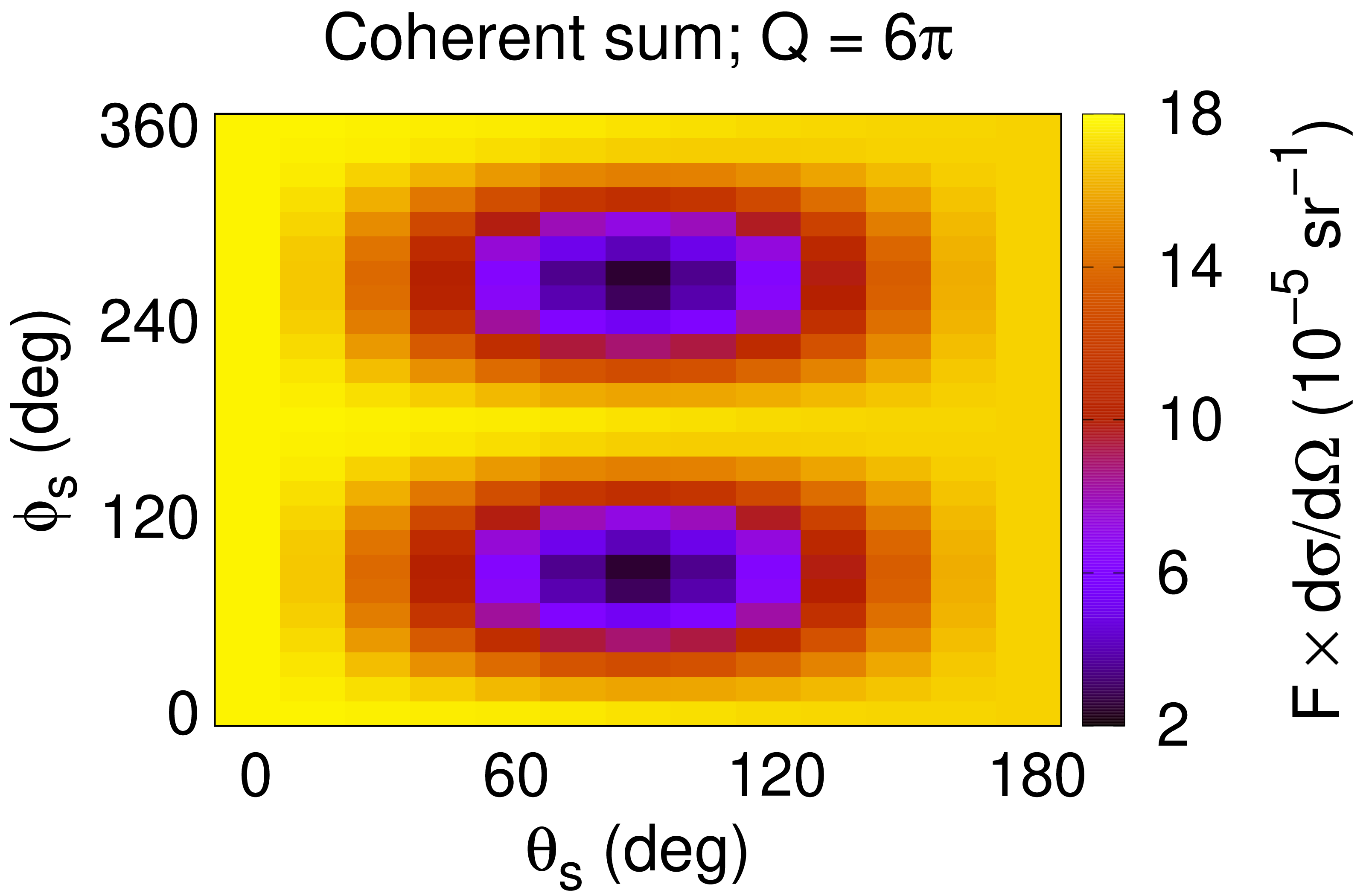}
\end{minipage}
\vspace{0.3cm}
\begin{minipage}[b][][b]{0.245\textwidth}
\includegraphics[width=\textwidth]{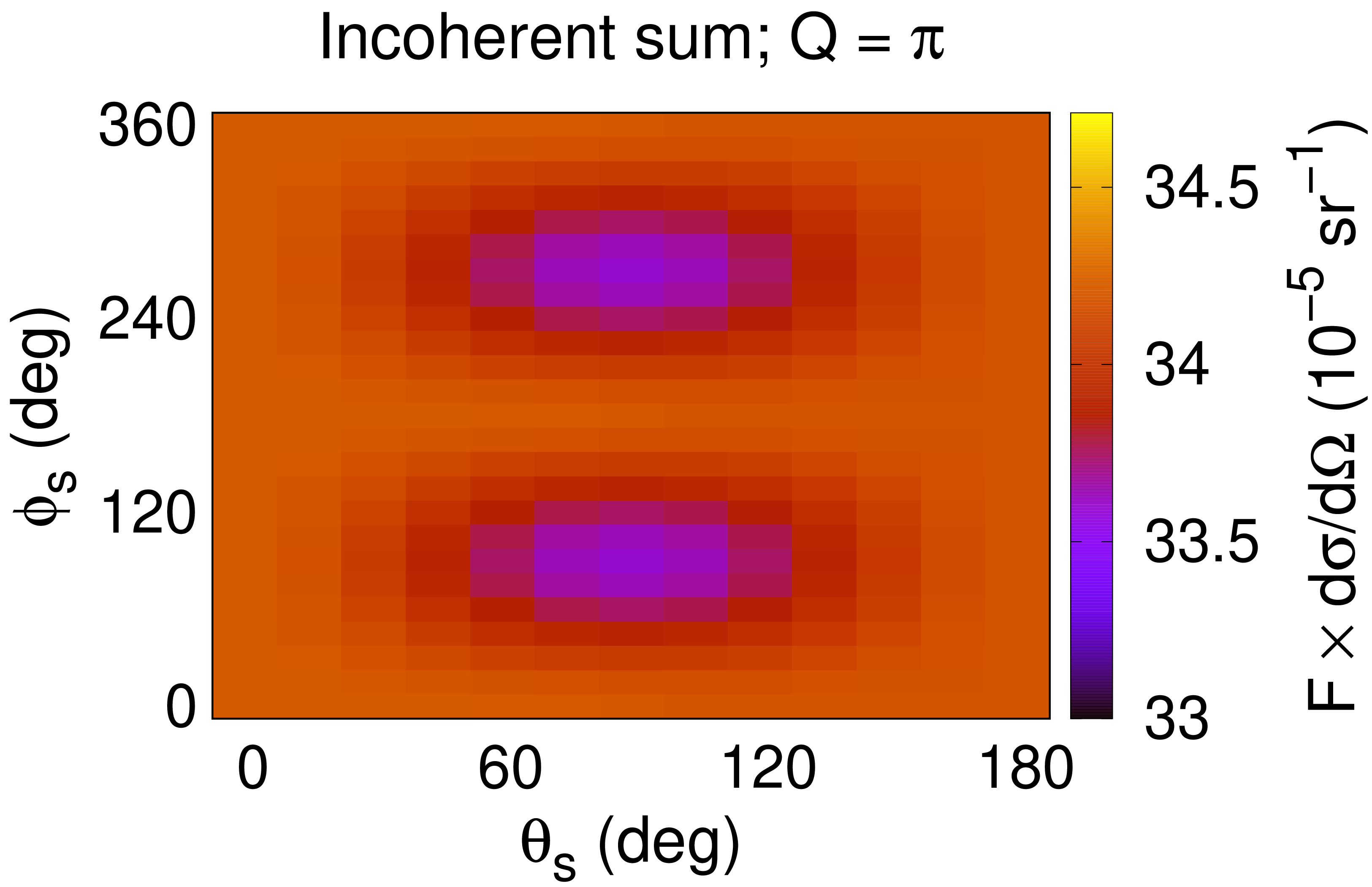}
\end{minipage}
\hspace{0.03cm}
\begin{minipage}[b][][b]{0.235\textwidth}
\includegraphics[width=\textwidth]{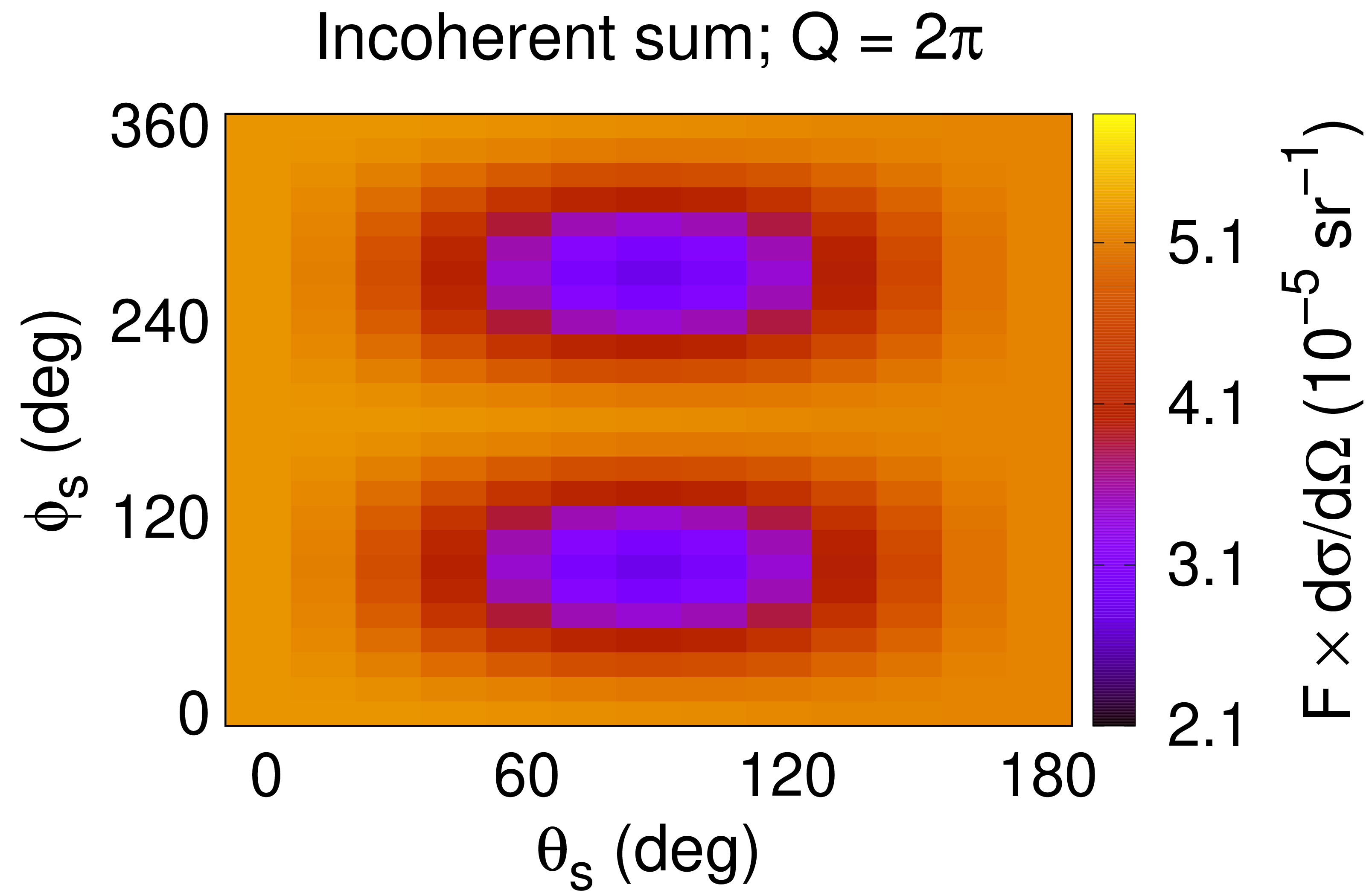}
\end{minipage}
\hspace{0.03cm}
\begin{minipage}[b][][b]{0.235\textwidth}
\includegraphics[width=\textwidth]{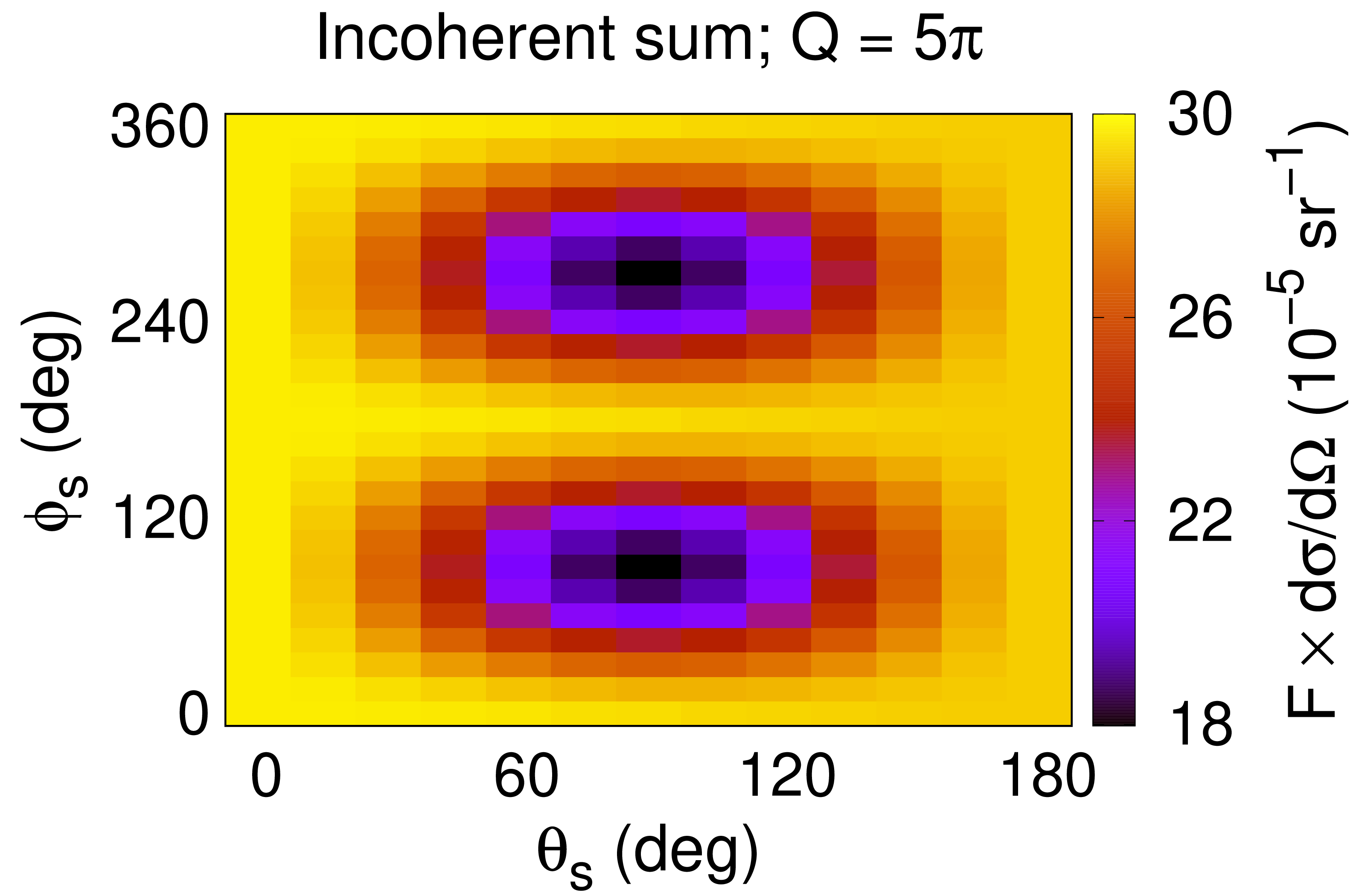}
\end{minipage}
\begin{minipage}[b][][b]{0.245\textwidth}
\includegraphics[width=\textwidth]{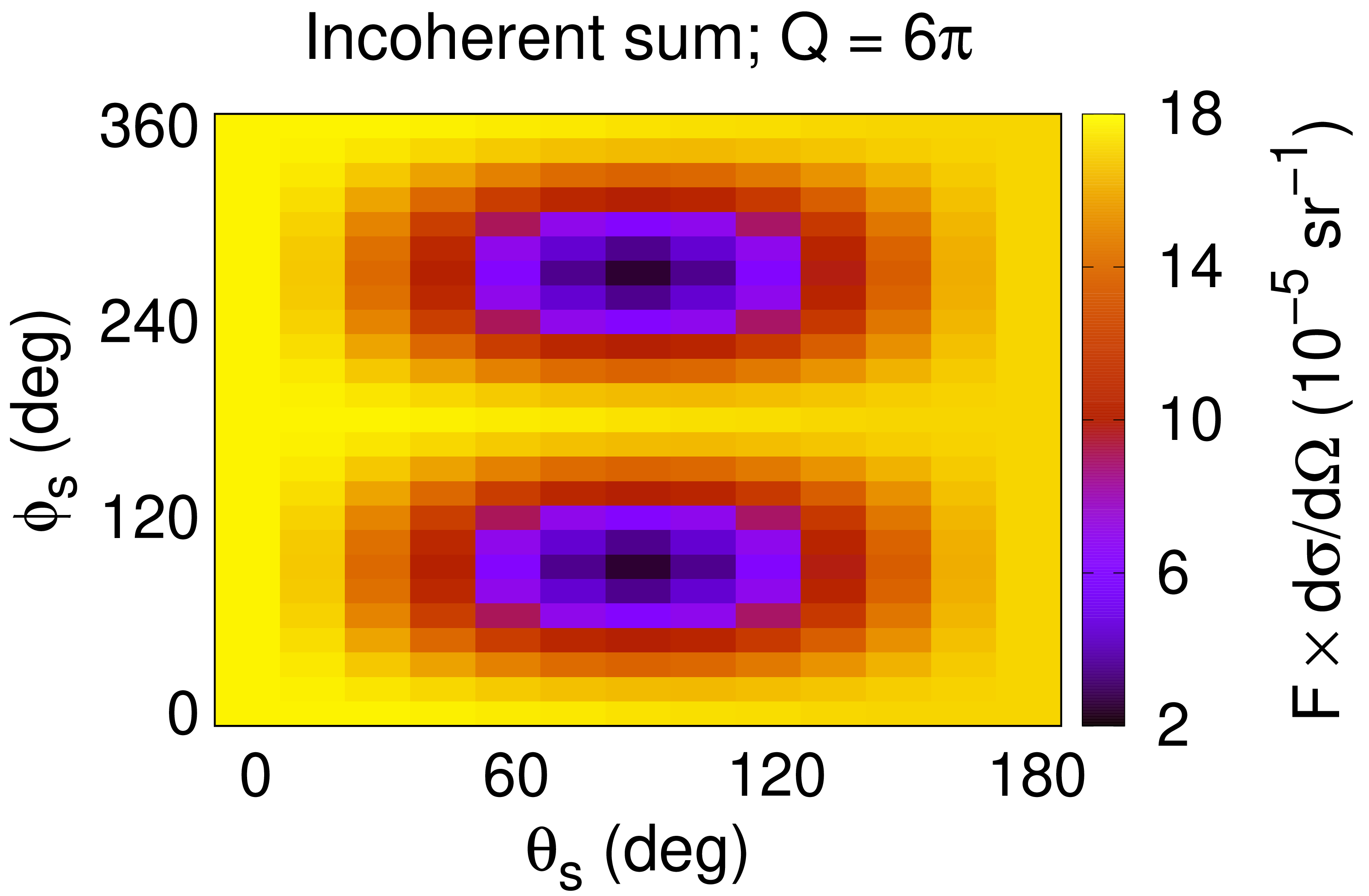}
\end{minipage}
\caption{ Angular distribution of photon yield for different pulse areas. The top and bottom row shows the results from a coherent and incoherent sum of the elastic scattering and resonant fluorescence channels. 
Here $t_{wid} = 0.25$ fs. Note that different scales are used for the color bars for different pulse areas. 
}
\label{Fig_angulardistrib_superposition}
\end{figure*}

The total photon yield from a single atom, calculated by integrating over all scattering angles, scattered photon energies, and summing over the outgoing photon polarizations for different incident pulse conditions, is shown in Figs.~\ref{Fig_signal_pulesarea}(b) and \ref{Fig_signal_pulesarea}(c). A scattered photon can result from either elastic scattering from any of the occupied 4-states in the system or from resonant fluorescence. During the pulse, if an outgoing photon mode (${\boldsymbol{k},\boldsymbol{\epsilon}}$) \textit{and} a final state of the scattered electron arise from both pathways, these pathways are indistinguishable and therefore interfere~\cite{Sakurai_adv}. The elastic scattering contribution to the scattering probability is linearly proportional to the incident intensity, while the resonant fluorescence contribution depends on the population of the excited state. We note that stimulated emission is not included in the scattering probability calculations as it cannot be described perturbatively in the outgoing field in this regime. However, its rate can be obtained from the Rabi amplitudes, and it produces outgoing photons in the same mode
of the incident field.

Fig.~\ref{Fig_signal_pulesarea}(b) reveals several trends. 
The $\pi$-type pulses exhibit the highest yields across different pulse durations, with their total photon yield for a single atom dominated by resonant fluorescence. These pulses cause the largest population transfer to the core-hole excited state resulting in maximum resonant fluorescence. Conversely, $2\pi$-type pulses have the smallest total photon yields across pulse durations. 
While the total photon yield of these pulses is substantially smaller than that of the $\pi$-type pulses, the elastic scattering contribution becomes comparable to the resonant fluorescence contribution, especially for shorter pulses.

For $\pi$-type pulses, the total photon yield decreases with increasing pulse area, but the opposite is true for 2$\pi$-type pulses. For $\pi$-type pulses, the maximum fluorescence occurs for Q = $\pi$ and increasing the pulse area further only serves to decrease this total photon yield due to increased photoionization. This is more pronounced for short pulses as they involve high intensities~(Fig.~\ref{Fig_signal_pulesarea}(a)). While increased photoionization from higher Q can reduce the scattered photon yield, there are two other competing effects for 2$\pi$-type pulses. First, higher Q results in higher resonant fluorescence during the pulse as the time-averaged population in the excited state increases. Second, unlike $\pi$-type pulses, the fraction of elastic scattering in the total photon yield is significant and the elastic scattering increases with Q. Thus, the total photon yield increases with Q in the explored regime (Fig.~\ref{Fig_signal_pulesarea}(b)).

Another trend in Fig.~\ref{Fig_signal_pulesarea}(b) is the decreasing difference in the total photon yield between $\pi$- and 2$\pi$-type pulses with increasing pulse duration. As the pulse duration increases, the amount of resonant fluorescence during the pulse increases. This reduces the difference between $\pi$- and 2$\pi$-type pulses. Similar trends were reported in the energy spectrum of resonant fluorescence \cite{Cavaletto_ResFluor_PRA}.

\begin{figure}
\resizebox{60mm}{!}{\includegraphics{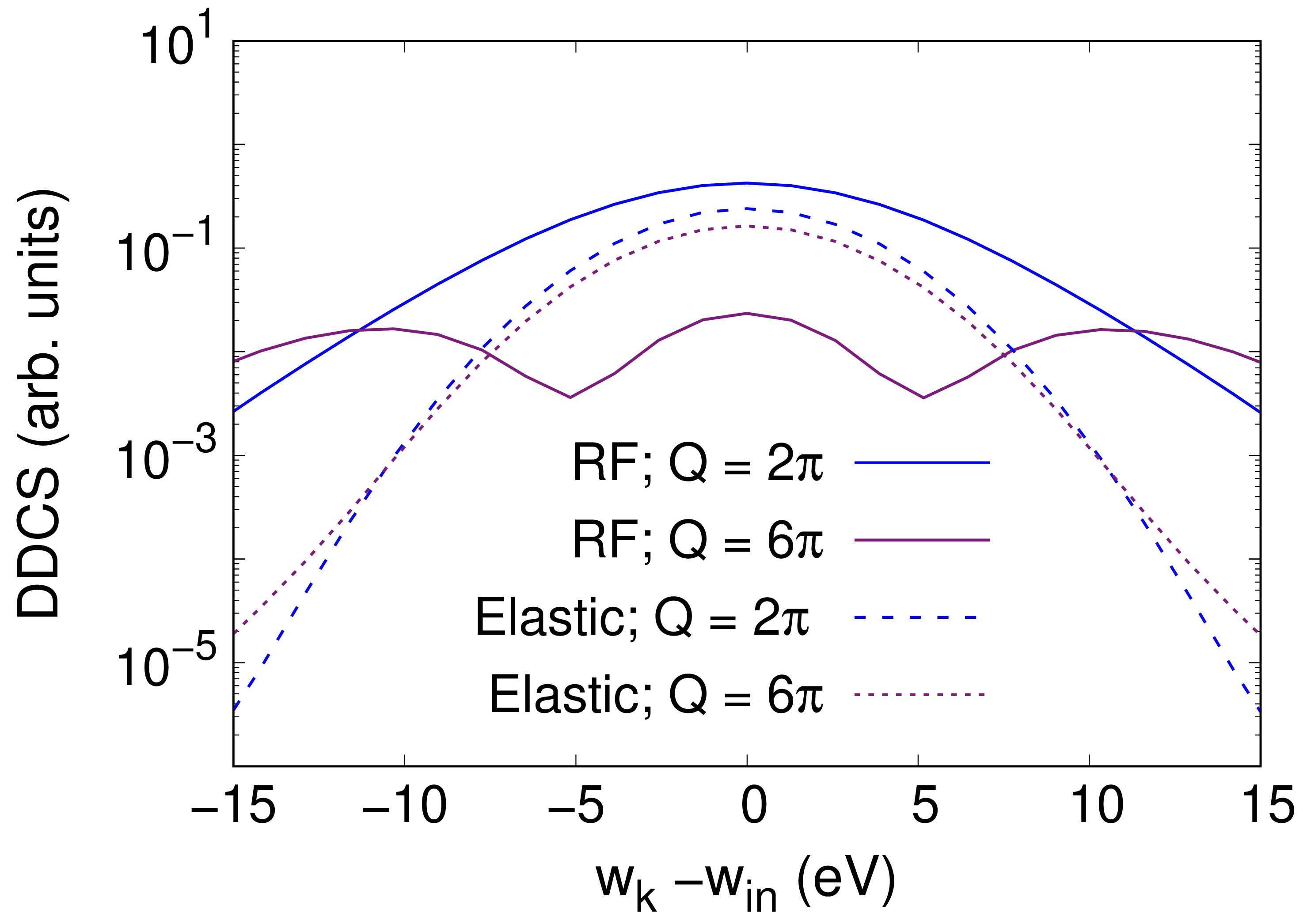}}
\caption{\label{DDCS_spectrum}
Comparison of a typical double differential cross section (DDCS) for different pulse areas from resonant fluorescence (RF) and elastic scattering. Here $\theta_s = 90\degree$ and $\phi_s = 45\degree$
}
\end{figure}
%
%
\begin{figure}
\begin{minipage}[t][3ex][t]{0.001\textwidth}
(a)
\end{minipage}
\begin{minipage}[t]{0.225\textwidth}
        \vspace{0cm}
        \includegraphics[width=\textwidth]{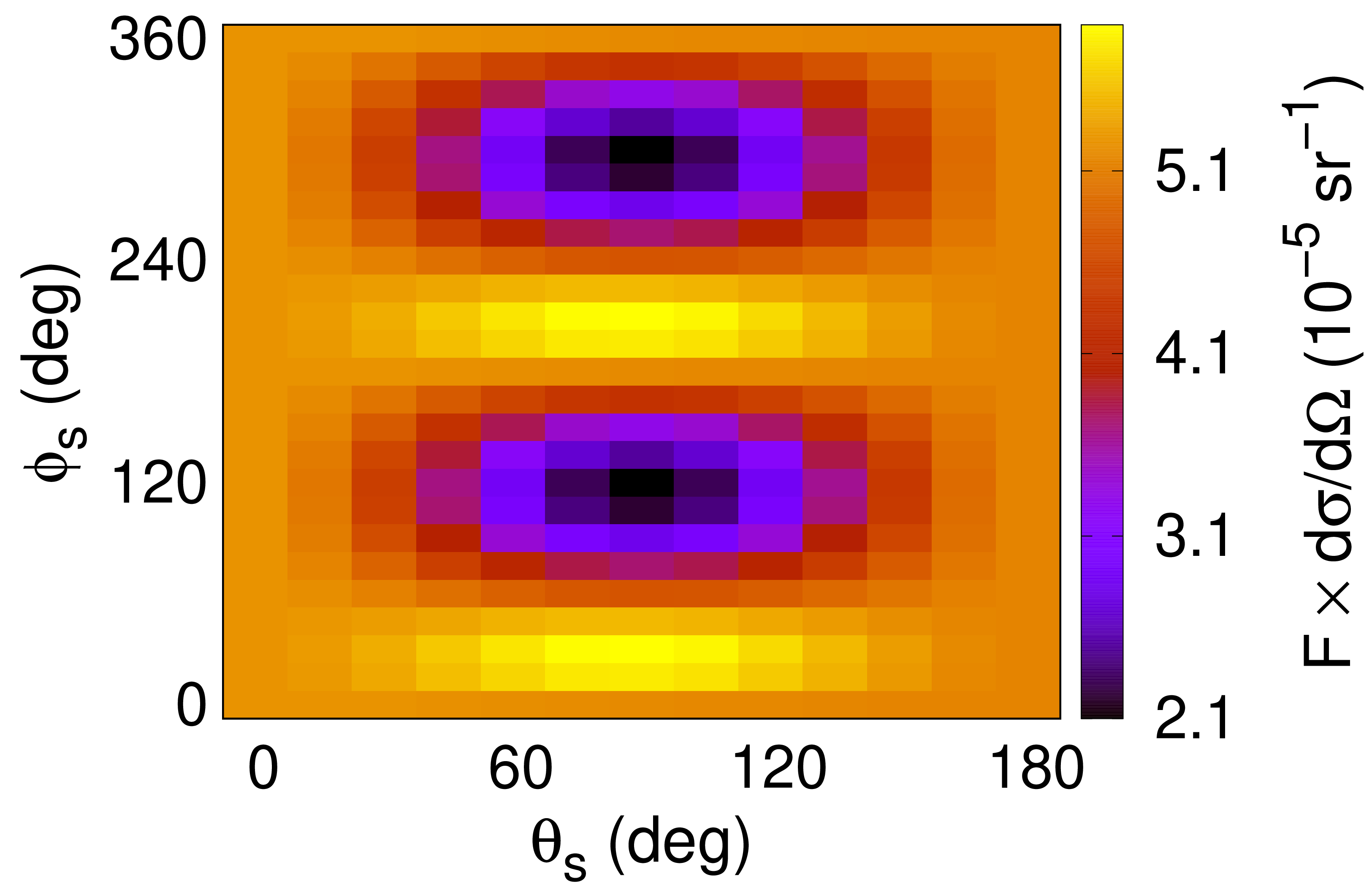}
\end{minipage}
\begin{minipage}[t][3ex][t]{0.001\textwidth}
          (b)
\end{minipage}
\begin{minipage}[t]{0.235\textwidth}
        \vspace{0cm}
        \includegraphics[width=\textwidth]{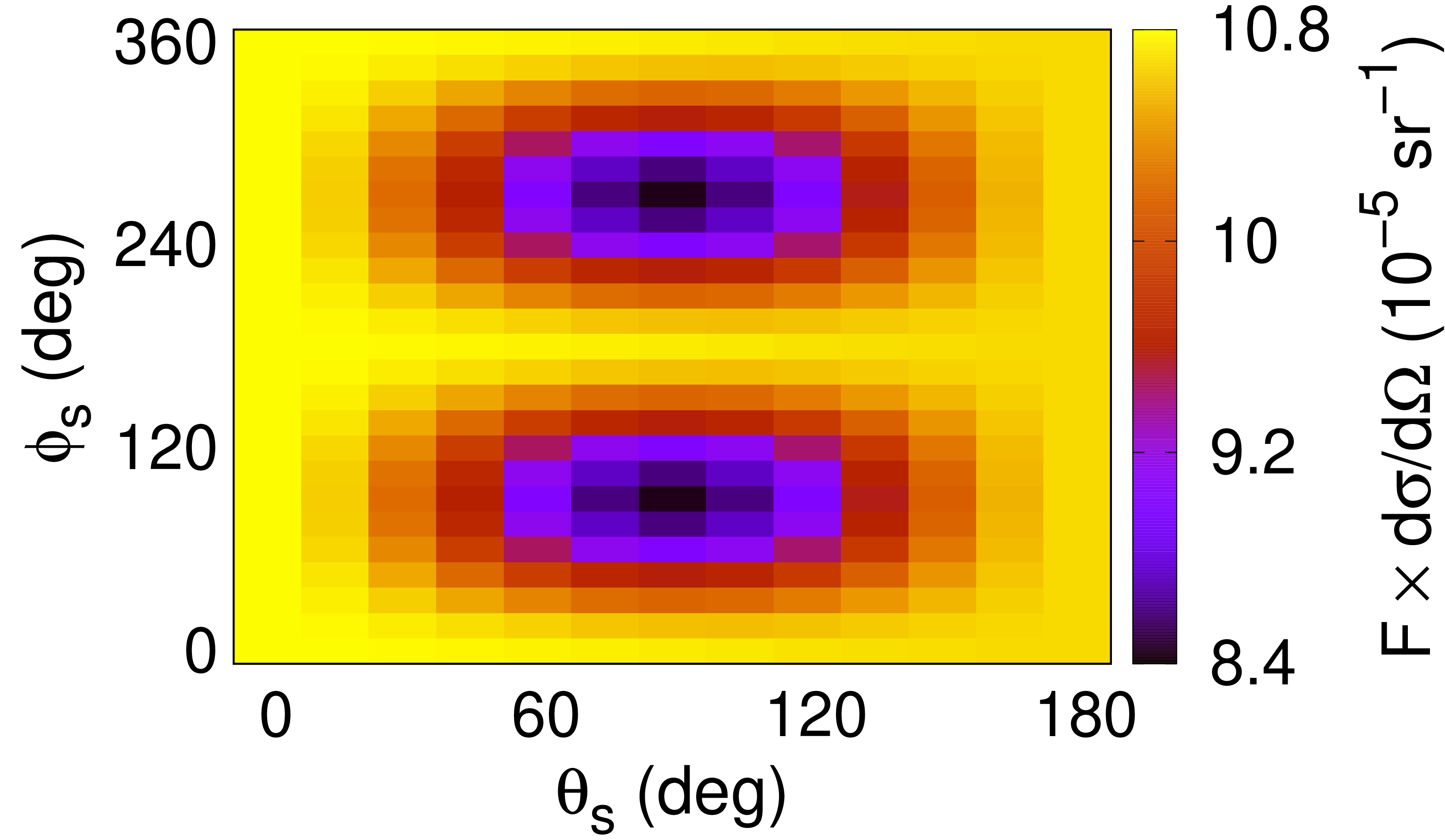}
\end{minipage}
\begin{minipage}[t][3ex][t]{0.001\textwidth}
        (c)
\end{minipage}
\begin{minipage}[t]{0.225\textwidth}
        \vspace{0cm}
        \includegraphics[width=\textwidth]{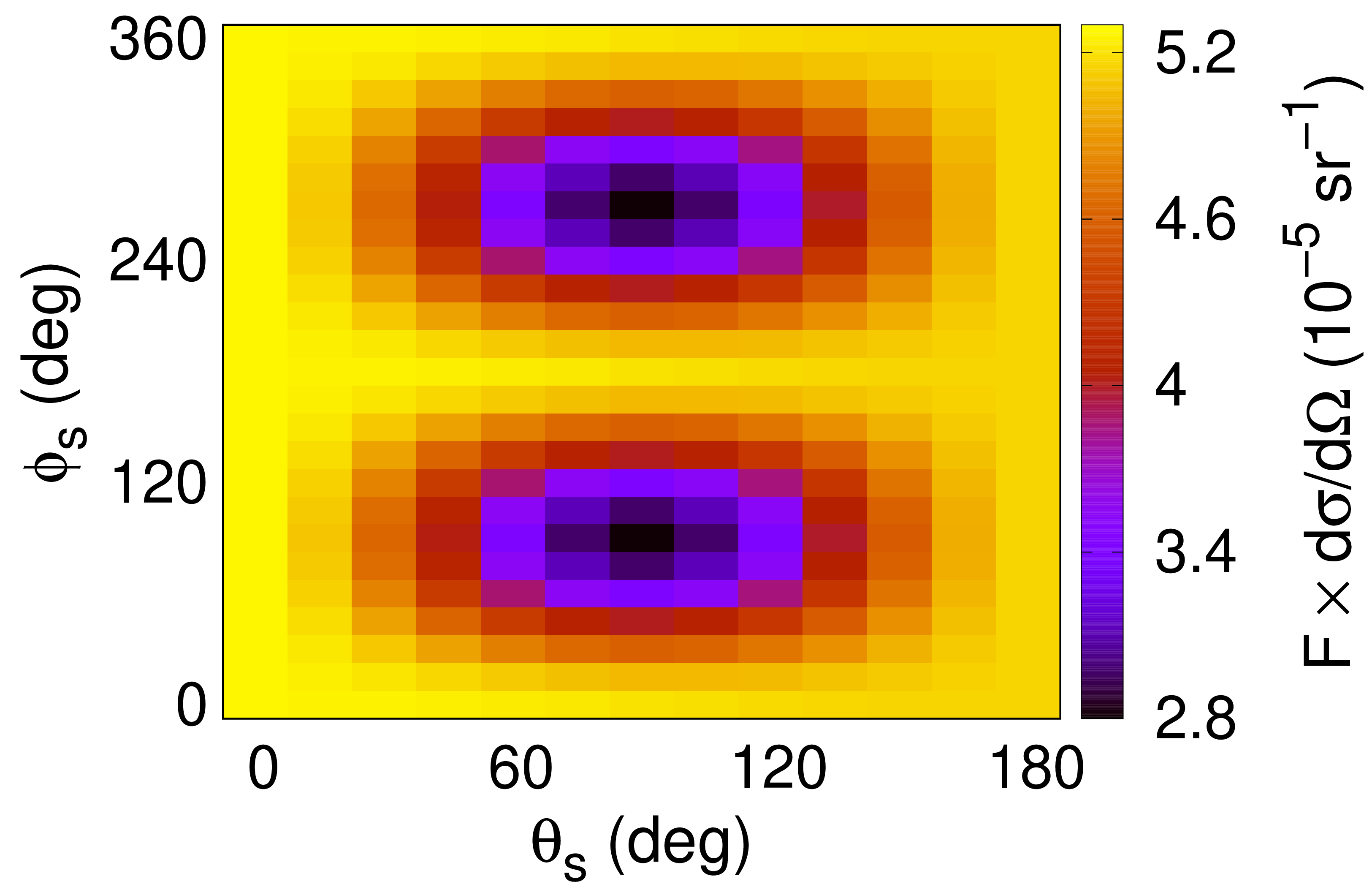}
\end{minipage}
\begin{minipage}[t][3ex][t]{0.001\textwidth}
       (d)
\end{minipage}
\begin{minipage}[t]{0.235\textwidth}
        \vspace{0cm}
        \includegraphics[width=\textwidth]{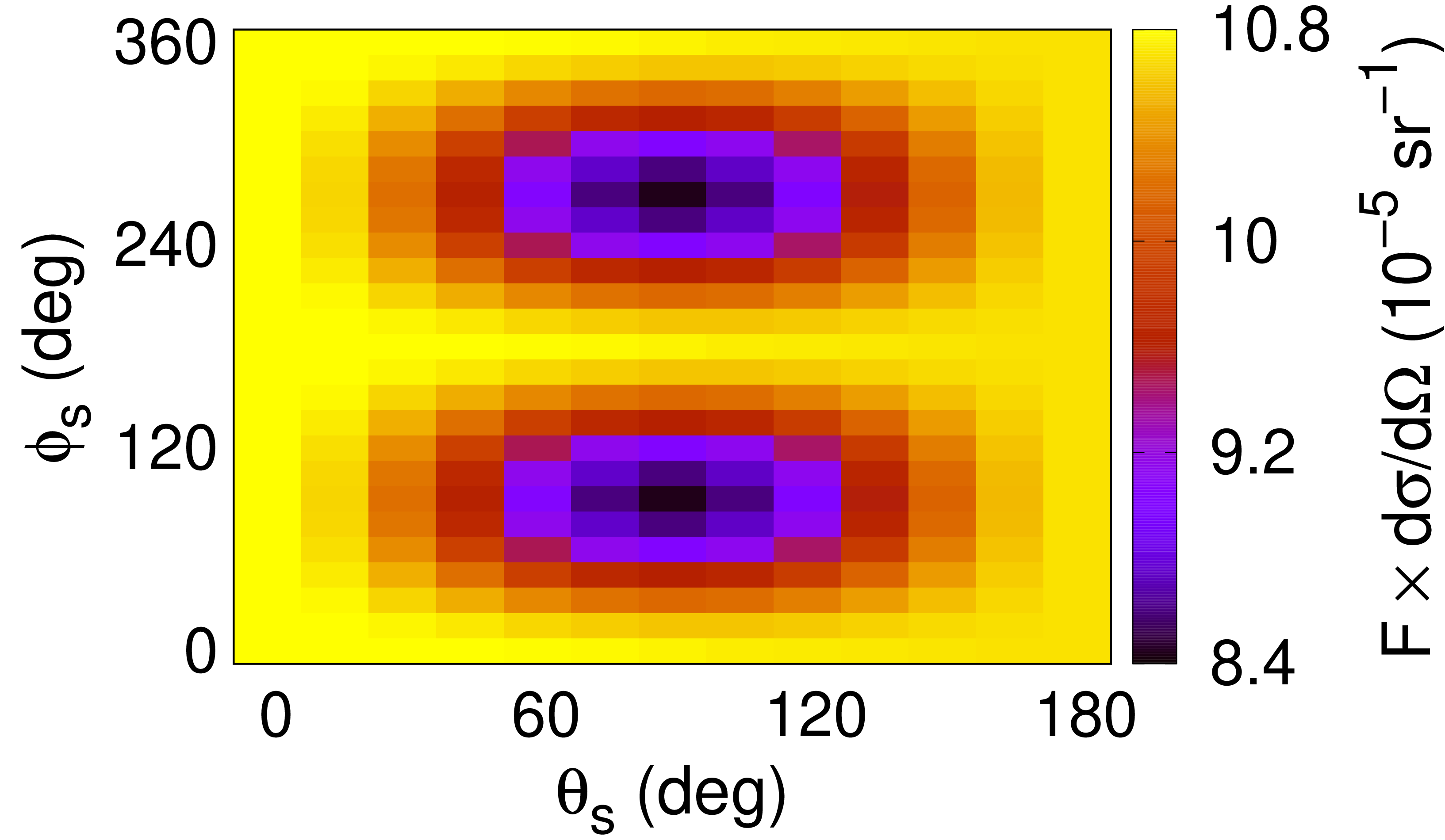}
\end{minipage}
\caption{ Angular distribution dependence on Ne\textsuperscript{+} initial state, $\ket{\psi_i}$. 
a) 
$\ket{\psi_i} = \frac{1}{\sqrt{3}} \big[ \ket{ 1s2p_{-1}^{-1} } - \ket{ 1s2p_0^{-1}} + \ket{ 1s2p_{+1}^{-1} } \big]$, (b),(d) $\ket{\psi_i} =  \ket{ 1s2p_0^{-1}}$, (c) $\ket{\psi_i} = \frac{1}{\sqrt{3}} \big[ -\ket{1s2p_{-1}^{-1} } + \ket{ 1s2p_0^{-1}} + \ket{ 1s2p_{+1}^{-1} } \big]$. Plots b) and d) denote the coherent and incoherent sum for the same initial state. Here $t_{wid} = 0.25$ fs and $Q = 2\pi$.
}
\label{Fig_initial_state_dep}
\end{figure}
More importantly, Fig.~\ref{Fig_signal_pulesarea}(c) shows that the yield of non-resonant x-ray scattering is at least one order of magnitude smaller than that the smallest yield of resonant x-ray scattering ($Q = 2\pi$). In non-resonant scattering,  the scattered photons come from the elastic scattering channel. Figure~\ref{Fig_signal_pulesarea}(c) also shows the estimated scattering yield from the Ne\textsuperscript{2+} population formed during the pulse which is not resonant with the incident pulse. Interestingly in this regime, the off-resonant yield from Ne\textsuperscript{+} is comparable to the yield from Ne\textsuperscript{2+}, which contributes to the background during resonant scattering from Ne\textsuperscript{+}. While the resonant x-ray scattering probability can have a complicated dependence on the charge density~\cite{Companion_article} and hence the structure, it offers site-specificity, which is absent in the non-resonant case.

Examining the angular distribution of the photon yield (Fig.~\ref{Fig_angulardistrib_superposition}), the effect of the interference between the elastic and resonant fluorescence channels is evident especially for small pulse areas. The total scattering yield, calculated from the sum of the amplitude of these channels, differs from the yield calculated using an incoherent sum of the individual channel probabilities.
The degree of interference between these 
channels decreases with increasing pulse area and effectively diminishes at high pulse areas.
For a given pulse duration, as the pulse intensity increases, a larger fraction of the resonant fluorescence spectrum during the pulse forms the sidebands and falls outside the bandwidth of the elastic scattering channel, which is similar to the incident pulse bandwidth, as shown in Fig.~\ref{DDCS_spectrum}. This reduces the amount of resonant fluorescence yield indistinguishable from the elastic scattering channel, thereby reducing the interference between these two channels.

Another observed trend is that the contrast or visibility of the interference fringes depends on the pulse intensity. The contrast can be computed as ($Y_{max}-Y_{min}$)/($Y_{max}+Y_{min}$), where $Y_{max}$ and $Y_{min}$ are the maximum and minimum photon yields of the angular distribution. For the scattering yield shown in Fig.~\ref{Fig_angulardistrib_superposition}, the contrasts for $\pi$, $2\pi$, $5\pi$ and $6\pi$ pulses are 0.02, 0.47, 0.24, and 0.75, respectively. The contrast stems only from the elastic channel and the interference term because the resonant fluorescence channel is effectively isotropic~\cite{Companion_article}. We find that the degree of contrast is typically dictated by the temporal overlap of the resonant fluorescence and elastic scattering channels. 2$\pi$-type pulses have a higher contrasts because the temporal emission window of both channels is restricted to the duration of the incident pulse. On the other hand, since the population is inverted at the end of a $\pi$-type pulse, the majority of fluorescence is emitted after the pulse, resulting in a smaller temporal overlap with the elastic scattering channel and less interference.

Our analysis shows that the angular distribution is shaped primarily by the interference between the elastic scattering amplitude (ESA) and the resonant fluorescence amplitude (RFA) in the final scattered states $1s2p_{\pm1}^{-1}$. More interestingly, there is no net interference in the angular distribution between the ESA and RFA in the final scattered state $1s2p_{0}^{-1}$ as their phases differ by $\pi/2$.  This phase difference can be understood from the fact the ESA and RFA are modulated by the probability amplitude of the ground state $1s2p_{0}^{-1}$ and core-excited state $1s^{-1}2p$ respectively, and the phases of these probability amplitudes differ by $\pi/2$.

Our results also suggest that the interference in the angular distribution is sensitive to the initial state of the Ne\textsuperscript{+}, $\ket{\psi_i}$. Fig.~\ref{Fig_initial_state_dep} shows that both the magnitude and phase of the three degenerate states in $\ket{\psi_i}$ can affect the scattering response.  
In the case that $\ket{\psi_i} =  \ket{ 1s2p_0^{-1}}$, Fig.~\ref{Fig_initial_state_dep}(b) shows that the angular distribution resembles that from the incoherent sum.  This lack of net interference is because the elastic scattering contribution from $1s2p_{\pm1}^{-1}$ is small due to its negligible population during the pulse, and there exists no net interference between the ESA and RFA in the final scattered state $1s2p_{0}^{-1}$. The resonant fluorescence yield is increased thrice in Fig.~\ref{Fig_initial_state_dep}(b)) relative to the superposition case because the entire initial state population can become excited.

For an initial superposition state, changing the sign of the $1s2p_{0}^{-1}$ state (see Fig.~\ref{Fig_initial_state_dep}(a) ) has the effect of imposing an extra phase of $e^{i\pi}$ on the resonant fluorescence channel while not altering the phase of the elastic scattered wave from the states $\ket{1s2p_{\pm1}^{-1}}$. This reverses the regions of constructive and destructive interferences. This is also equivalent to changing the signs of weights of both $\ket{1s2p_{-1}^{-1}}$ and $\ket{1s2p_{1}^{-1}}$. Finally, changing the sign for either $\ket{1s2p_{-1}^{-1}}$ or $\ket{1s2p_{+1}^{-1}}$ results in opposite kinds of interference from these states, giving rise to no overall interference.

In summary, we investigate single-atom response of resonant ultrafast scattering in intense, few-femtosecond and attosecond pulses. The effect of x-ray-driven Rabi oscillations on the total photon yield is presented. The scattered photon yield from transient resonance is found to strongly depend on the pulse area, deviating from the linear scattering model. Instead of increasing with pulse intensity, it exhibits an upper bound. Compared to non-resonant scattering conditions, resonant scatterings are at least an order of magnitude higher, suggesting that resonant scattering can be exploited for high-resolution imaging with elemental contrast. 
The contrast and appearance of fringes in the angular distribution are sensitive to the degree of interference between the elastic and resonant fluorescence channels. For small pulse areas, the interference is sensitive to the initial state of the transient resonance, while for large pulse areas, the signature of interference between these channels diminishes.  Thus, our study shows that pulse area can be used to control temporal emission profile, angular distribution and yield of the scattered photon.  Most importantly, it can be exploited to optimize scattering signals that carry structural information and improve contrast within the image. In the future, extending our approach for multi-atom systems and in hard x-ray regime will allow investigations of the potential of using resonant x-ray scattering for imaging the structure of heterogeneous systems \cite{Bhowmick-IUCRJ-2023}.

We are grateful to the members of the Atomic, Molecular and Optical Physics group at Argonne National Laboratory for fruitful discussions and also to T. Gorkhover, A. Ulmer, S. Kuschel, R. Radloff for discussions on their Neon cluster experiment. A.V. thanks F. Robicheaux for suggestions. This work was supported by the U.S. Department of Energy, Office of Basic Energy Sciences, Division of Chemical Sciences, Geosciences, and Biosciences through Argonne National Laboratory. Argonne is a U.S. Department of Energy laboratory managed by UChicago Argonne, LLC, under Contract No. DE-AC02-06CH11357. We gratefully acknowledge the computing resources provided on Improv, a high-performance computing cluster operated by the Laboratory Computing Resource Center at Argonne National Laboratory.

\bibliography{References.bib}

\begin{thebibliography}{43}%
\makeatletter
\providecommand \@ifxundefined [1]{%
 \@ifx{#1\undefined}
}%
\providecommand \@ifnum [1]{%
 \ifnum #1\expandafter \@firstoftwo
 \else \expandafter \@secondoftwo
 \fi
}%
\providecommand \@ifx [1]{%
 \ifx #1\expandafter \@firstoftwo
 \else \expandafter \@secondoftwo
 \fi
}%
\providecommand \natexlab [1]{#1}%
\providecommand \enquote  [1]{``#1''}%
\providecommand \bibnamefont  [1]{#1}%
\providecommand \bibfnamefont [1]{#1}%
\providecommand \citenamefont [1]{#1}%
\providecommand \href@noop [0]{\@secondoftwo}%
\providecommand \href [0]{\begingroup \@sanitize@url \@href}%
\providecommand \@href[1]{\@@startlink{#1}\@@href}%
\providecommand \@@href[1]{\endgroup#1\@@endlink}%
\providecommand \@sanitize@url [0]{\catcode `\\12\catcode `\$12\catcode `\&12\catcode `\#12\catcode `\^12\catcode `\_12\catcode `\%12\relax}%
\providecommand \@@startlink[1]{}%
\providecommand \@@endlink[0]{}%
\providecommand \url  [0]{\begingroup\@sanitize@url \@url }%
\providecommand \@url [1]{\endgroup\@href {#1}{\urlprefix }}%
\providecommand \urlprefix  [0]{URL }%
\providecommand \Eprint [0]{\href }%
\providecommand \doibase [0]{https://doi.org/}%
\providecommand \selectlanguage [0]{\@gobble}%
\providecommand \bibinfo  [0]{\@secondoftwo}%
\providecommand \bibfield  [0]{\@secondoftwo}%
\providecommand \translation [1]{[#1]}%
\providecommand \BibitemOpen [0]{}%
\providecommand \bibitemStop [0]{}%
\providecommand \bibitemNoStop [0]{.\EOS\space}%
\providecommand \EOS [0]{\spacefactor3000\relax}%
\providecommand \BibitemShut  [1]{\csname bibitem#1\endcsname}%
\let\auto@bib@innerbib\@empty
\bibitem [{\citenamefont {Emma}\ \emph {et~al.}(2010)\citenamefont {Emma}, \citenamefont {Akre}, \citenamefont {Arthur}, \citenamefont {Bionta}, \citenamefont {Bostedt}, \citenamefont {Bozek}, \citenamefont {Brachmann}, \citenamefont {Bucksbaum}, \citenamefont {Coffee}, \citenamefont {Decker} \emph {et~al.}}]{XFEL1}%
  \BibitemOpen
  \bibfield  {author} {\bibinfo {author} {\bibfnamefont {P.}~\bibnamefont {Emma}}, \bibinfo {author} {\bibfnamefont {R.}~\bibnamefont {Akre}}, \bibinfo {author} {\bibfnamefont {J.}~\bibnamefont {Arthur}}, \bibinfo {author} {\bibfnamefont {R.}~\bibnamefont {Bionta}}, \bibinfo {author} {\bibfnamefont {C.}~\bibnamefont {Bostedt}}, \bibinfo {author} {\bibfnamefont {J.}~\bibnamefont {Bozek}}, \bibinfo {author} {\bibfnamefont {A.}~\bibnamefont {Brachmann}}, \bibinfo {author} {\bibfnamefont {P.}~\bibnamefont {Bucksbaum}}, \bibinfo {author} {\bibfnamefont {R.}~\bibnamefont {Coffee}}, \bibinfo {author} {\bibfnamefont {F.-J.}\ \bibnamefont {Decker}}, \emph {et~al.},\ }\bibfield  {title} {\bibinfo {title} {First lasing and operation of an {\aa}ngstrom-wavelength free-electron laser},\ }\href@noop {} {\bibfield  {journal} {\bibinfo  {journal} {Nat. Photon.}\ }\textbf {\bibinfo {volume} {4}},\ \bibinfo {pages} {641} (\bibinfo {year} {2010})}\BibitemShut {NoStop}%
\bibitem [{\citenamefont {Altarelli}(2011)}]{EuropeanXFEL1}%
  \BibitemOpen
  \bibfield  {author} {\bibinfo {author} {\bibfnamefont {M.}~\bibnamefont {Altarelli}},\ }\bibfield  {title} {\bibinfo {title} {The european x-ray free-electron laser facility in hamburg},\ }\href {https://doi.org/https://doi.org/10.1016/j.nimb.2011.04.034} {\bibfield  {journal} {\bibinfo  {journal} {Nuclear Instruments and Methods in Physics Research Section B: Beam Interactions with Materials and Atoms}\ }\textbf {\bibinfo {volume} {269}},\ \bibinfo {pages} {2845 } (\bibinfo {year} {2011})},\ \bibinfo {note} {proceedings of the 10th European Conference on Accelerators in Applied Research and Technology (ECAART10)}\BibitemShut {NoStop}%
\bibitem [{\citenamefont {Ishikawa}\ \emph {et~al.}(2012)\citenamefont {Ishikawa}, \citenamefont {Aoyagi}, \citenamefont {Asaka}, \citenamefont {Asano}, \citenamefont {Azumi}, \citenamefont {Bizen}, \citenamefont {Ego}, \citenamefont {Fukami}, \citenamefont {Fukui}, \citenamefont {Furukawa} \emph {et~al.}}]{SACLA_1}%
  \BibitemOpen
  \bibfield  {author} {\bibinfo {author} {\bibfnamefont {T.}~\bibnamefont {Ishikawa}}, \bibinfo {author} {\bibfnamefont {H.}~\bibnamefont {Aoyagi}}, \bibinfo {author} {\bibfnamefont {T.}~\bibnamefont {Asaka}}, \bibinfo {author} {\bibfnamefont {Y.}~\bibnamefont {Asano}}, \bibinfo {author} {\bibfnamefont {N.}~\bibnamefont {Azumi}}, \bibinfo {author} {\bibfnamefont {T.}~\bibnamefont {Bizen}}, \bibinfo {author} {\bibfnamefont {H.}~\bibnamefont {Ego}}, \bibinfo {author} {\bibfnamefont {K.}~\bibnamefont {Fukami}}, \bibinfo {author} {\bibfnamefont {T.}~\bibnamefont {Fukui}}, \bibinfo {author} {\bibfnamefont {Y.}~\bibnamefont {Furukawa}}, \emph {et~al.},\ }\bibfield  {title} {\bibinfo {title} {A compact {X}-ray free-electron laser emitting in the sub-{\aa}ngstr{\"o}m region},\ }\href@noop {} {\bibfield  {journal} {\bibinfo  {journal} {Nat. Photon.}\ }\textbf {\bibinfo {volume} {6}},\ \bibinfo {pages} {540} (\bibinfo {year} {2012})}\BibitemShut {NoStop}%
\bibitem [{\citenamefont {Milne}\ \emph {et~al.}(2017)\citenamefont {Milne}, \citenamefont {Schietinger}, \citenamefont {Aiba}, \citenamefont {Alarcon}, \citenamefont {Alex}, \citenamefont {Anghel}, \citenamefont {Arsov}, \citenamefont {Beard}, \citenamefont {Beaud}, \citenamefont {Bettoni} \emph {et~al.}}]{swissFEL_1}%
  \BibitemOpen
  \bibfield  {author} {\bibinfo {author} {\bibfnamefont {C.~J.}\ \bibnamefont {Milne}}, \bibinfo {author} {\bibfnamefont {T.}~\bibnamefont {Schietinger}}, \bibinfo {author} {\bibfnamefont {M.}~\bibnamefont {Aiba}}, \bibinfo {author} {\bibfnamefont {A.}~\bibnamefont {Alarcon}}, \bibinfo {author} {\bibfnamefont {J.}~\bibnamefont {Alex}}, \bibinfo {author} {\bibfnamefont {A.}~\bibnamefont {Anghel}}, \bibinfo {author} {\bibfnamefont {V.}~\bibnamefont {Arsov}}, \bibinfo {author} {\bibfnamefont {C.}~\bibnamefont {Beard}}, \bibinfo {author} {\bibfnamefont {P.}~\bibnamefont {Beaud}}, \bibinfo {author} {\bibfnamefont {S.}~\bibnamefont {Bettoni}}, \emph {et~al.},\ }\bibfield  {title} {\bibinfo {title} {Swissfel: the swiss x-ray free electron laser},\ }\href@noop {} {\bibfield  {journal} {\bibinfo  {journal} {Applied Sciences}\ }\textbf {\bibinfo {volume} {7}},\ \bibinfo {pages} {720} (\bibinfo {year} {2017})}\BibitemShut {NoStop}%
\bibitem [{\citenamefont {Lindroth}\ \emph {et~al.}(2019)\citenamefont {Lindroth}, \citenamefont {Calegari}, \citenamefont {Young}, \citenamefont {Harmand}, \citenamefont {Dudovich}, \citenamefont {Berrah},\ and\ \citenamefont {Smirnova}}]{XFELscience_overview_Young}%
  \BibitemOpen
  \bibfield  {author} {\bibinfo {author} {\bibfnamefont {E.}~\bibnamefont {Lindroth}}, \bibinfo {author} {\bibfnamefont {F.}~\bibnamefont {Calegari}}, \bibinfo {author} {\bibfnamefont {L.}~\bibnamefont {Young}}, \bibinfo {author} {\bibfnamefont {M.}~\bibnamefont {Harmand}}, \bibinfo {author} {\bibfnamefont {N.}~\bibnamefont {Dudovich}}, \bibinfo {author} {\bibfnamefont {N.}~\bibnamefont {Berrah}},\ and\ \bibinfo {author} {\bibfnamefont {O.}~\bibnamefont {Smirnova}},\ }\bibfield  {title} {\bibinfo {title} {Challenges and opportunities in attosecond and xfel science},\ }\href@noop {} {\bibfield  {journal} {\bibinfo  {journal} {Nature Reviews Physics}\ }\textbf {\bibinfo {volume} {1}},\ \bibinfo {pages} {107} (\bibinfo {year} {2019})}\BibitemShut {NoStop}%
\bibitem [{\citenamefont {Pellegrini}(2020)}]{XFELdevelopmentsummary_2020}%
  \BibitemOpen
  \bibfield  {author} {\bibinfo {author} {\bibfnamefont {C.}~\bibnamefont {Pellegrini}},\ }\bibfield  {title} {\bibinfo {title} {The development of xfels},\ }\href@noop {} {\bibfield  {journal} {\bibinfo  {journal} {Nature Reviews Physics}\ }\textbf {\bibinfo {volume} {2}},\ \bibinfo {pages} {330} (\bibinfo {year} {2020})}\BibitemShut {NoStop}%
\bibitem [{\citenamefont {Neutze}\ \emph {et~al.}(2000)\citenamefont {Neutze}, \citenamefont {Wouts}, \citenamefont {Van~der Spoel}, \citenamefont {Weckert},\ and\ \citenamefont {Hajdu}}]{neutze2000_imagebeforedestroy}%
  \BibitemOpen
  \bibfield  {author} {\bibinfo {author} {\bibfnamefont {R.}~\bibnamefont {Neutze}}, \bibinfo {author} {\bibfnamefont {R.}~\bibnamefont {Wouts}}, \bibinfo {author} {\bibfnamefont {D.}~\bibnamefont {Van~der Spoel}}, \bibinfo {author} {\bibfnamefont {E.}~\bibnamefont {Weckert}},\ and\ \bibinfo {author} {\bibfnamefont {J.}~\bibnamefont {Hajdu}},\ }\bibfield  {title} {\bibinfo {title} {Potential for biomolecular imaging with femtosecond x-ray pulses},\ }\href@noop {} {\bibfield  {journal} {\bibinfo  {journal} {Nature}\ }\textbf {\bibinfo {volume} {406}},\ \bibinfo {pages} {752} (\bibinfo {year} {2000})}\BibitemShut {NoStop}%
\bibitem [{\citenamefont {Kirz}\ \emph {et~al.}(1995)\citenamefont {Kirz}, \citenamefont {Jacobsen},\ and\ \citenamefont {Howells}}]{Kirz_Jacobsen_Howells_1995}%
  \BibitemOpen
  \bibfield  {author} {\bibinfo {author} {\bibfnamefont {J.}~\bibnamefont {Kirz}}, \bibinfo {author} {\bibfnamefont {C.}~\bibnamefont {Jacobsen}},\ and\ \bibinfo {author} {\bibfnamefont {M.}~\bibnamefont {Howells}},\ }\bibfield  {title} {\bibinfo {title} {Soft x-ray microscopes and their biological applications},\ }\href {https://doi.org/10.1017/S0033583500003139} {\bibfield  {journal} {\bibinfo  {journal} {Quarterly Reviews of Biophysics}\ }\textbf {\bibinfo {volume} {28}},\ \bibinfo {pages} {33–130} (\bibinfo {year} {1995})}\BibitemShut {NoStop}%
\bibitem [{\citenamefont {Young}\ \emph {et~al.}(2010)\citenamefont {Young}, \citenamefont {Kanter}, \citenamefont {Kr{\"a}ssig}, \citenamefont {Li}, \citenamefont {March}, \citenamefont {Pratt}, \citenamefont {Santra}, \citenamefont {Southworth}, \citenamefont {Rohringer}, \citenamefont {DiMauro}, \citenamefont {Doumy}, \citenamefont {Roedig}, \citenamefont {Berrah}, \citenamefont {Fang}, \citenamefont {Hoener}, \citenamefont {Bucksbaum}, \citenamefont {Cryan}, \citenamefont {Ghimire}, \citenamefont {Glownia}, \citenamefont {Reis}, \citenamefont {Bozek}, \citenamefont {Bostedt},\ and\ \citenamefont {Messerschmidt}}]{Young2010}%
  \BibitemOpen
  \bibfield  {author} {\bibinfo {author} {\bibfnamefont {L.}~\bibnamefont {Young}}, \bibinfo {author} {\bibfnamefont {E.~P.}\ \bibnamefont {Kanter}}, \bibinfo {author} {\bibfnamefont {B.}~\bibnamefont {Kr{\"a}ssig}}, \bibinfo {author} {\bibfnamefont {Y.}~\bibnamefont {Li}}, \bibinfo {author} {\bibfnamefont {A.~M.}\ \bibnamefont {March}}, \bibinfo {author} {\bibfnamefont {S.~T.}\ \bibnamefont {Pratt}}, \bibinfo {author} {\bibfnamefont {R.}~\bibnamefont {Santra}}, \bibinfo {author} {\bibfnamefont {S.~H.}\ \bibnamefont {Southworth}}, \bibinfo {author} {\bibfnamefont {N.}~\bibnamefont {Rohringer}}, \bibinfo {author} {\bibfnamefont {L.~F.}\ \bibnamefont {DiMauro}}, \bibinfo {author} {\bibfnamefont {G.}~\bibnamefont {Doumy}}, \bibinfo {author} {\bibfnamefont {C.~A.}\ \bibnamefont {Roedig}}, \bibinfo {author} {\bibfnamefont {N.}~\bibnamefont {Berrah}}, \bibinfo {author} {\bibfnamefont {L.}~\bibnamefont {Fang}}, \bibinfo {author} {\bibfnamefont {M.}~\bibnamefont {Hoener}}, \bibinfo {author} {\bibfnamefont {P.~H.}\
  \bibnamefont {Bucksbaum}}, \bibinfo {author} {\bibfnamefont {J.~P.}\ \bibnamefont {Cryan}}, \bibinfo {author} {\bibfnamefont {S.}~\bibnamefont {Ghimire}}, \bibinfo {author} {\bibfnamefont {J.~M.}\ \bibnamefont {Glownia}}, \bibinfo {author} {\bibfnamefont {D.~A.}\ \bibnamefont {Reis}}, \bibinfo {author} {\bibfnamefont {J.~D.}\ \bibnamefont {Bozek}}, \bibinfo {author} {\bibfnamefont {C.}~\bibnamefont {Bostedt}},\ and\ \bibinfo {author} {\bibfnamefont {M.}~\bibnamefont {Messerschmidt}},\ }\bibfield  {title} {\bibinfo {title} {Femtosecond electronic response of atoms to ultra-intense x-rays},\ }\href {https://doi.org/10.1038/nature09177} {\bibfield  {journal} {\bibinfo  {journal} {Nature}\ }\textbf {\bibinfo {volume} {466}},\ \bibinfo {pages} {56} (\bibinfo {year} {2010})}\BibitemShut {NoStop}%
\bibitem [{\citenamefont {Rudek}\ \emph {et~al.}(2012)\citenamefont {Rudek}, \citenamefont {Son}, \citenamefont {Foucar}, \citenamefont {Epp}, \citenamefont {Erk}, \citenamefont {Hartmann}, \citenamefont {Adolph}, \citenamefont {Andritschke}, \citenamefont {Aquila}, \citenamefont {Berrah}, \citenamefont {Bostedt}, \citenamefont {Bozek}, \citenamefont {Coppola}, \citenamefont {Filsinger}, \citenamefont {Gorke}, \citenamefont {Gorkhover}, \citenamefont {Graafsma}, \citenamefont {Gumprecht}, \citenamefont {Hartmann}, \citenamefont {Hauser}, \citenamefont {Herrmann}, \citenamefont {Hirsemann}, \citenamefont {Holl}, \citenamefont {Hoemke}, \citenamefont {Journel}, \citenamefont {Kaiser}, \citenamefont {Kimmel}, \citenamefont {Krasniqi}, \citenamefont {Kuehnel}, \citenamefont {Matysek}, \citenamefont {Messerschmidt}, \citenamefont {Miesner}, \citenamefont {Moeller}, \citenamefont {Moshammer}, \citenamefont {Nagaya}, \citenamefont {Nilsson}, \citenamefont {Potdevin}, \citenamefont {Pietschner}, \citenamefont
  {Reich}, \citenamefont {Rupp}, \citenamefont {Schaller}, \citenamefont {Schlichting}, \citenamefont {Schmidt}, \citenamefont {Schopper}, \citenamefont {Schorb}, \citenamefont {Schroeter}, \citenamefont {Schulz}, \citenamefont {Simon}, \citenamefont {Soltau}, \citenamefont {Strueder}, \citenamefont {Ueda}, \citenamefont {Weidenspointner}, \citenamefont {Santra}, \citenamefont {Ullrich}, \citenamefont {Rudenko},\ and\ \citenamefont {Rolles}}]{Rudek-2012-NatPho}%
  \BibitemOpen
  \bibfield  {author} {\bibinfo {author} {\bibfnamefont {B.}~\bibnamefont {Rudek}}, \bibinfo {author} {\bibfnamefont {S.-K.}\ \bibnamefont {Son}}, \bibinfo {author} {\bibfnamefont {L.}~\bibnamefont {Foucar}}, \bibinfo {author} {\bibfnamefont {S.~W.}\ \bibnamefont {Epp}}, \bibinfo {author} {\bibfnamefont {B.}~\bibnamefont {Erk}}, \bibinfo {author} {\bibfnamefont {R.}~\bibnamefont {Hartmann}}, \bibinfo {author} {\bibfnamefont {M.}~\bibnamefont {Adolph}}, \bibinfo {author} {\bibfnamefont {R.}~\bibnamefont {Andritschke}}, \bibinfo {author} {\bibfnamefont {A.}~\bibnamefont {Aquila}}, \bibinfo {author} {\bibfnamefont {N.}~\bibnamefont {Berrah}}, \bibinfo {author} {\bibfnamefont {C.}~\bibnamefont {Bostedt}}, \bibinfo {author} {\bibfnamefont {J.}~\bibnamefont {Bozek}}, \bibinfo {author} {\bibfnamefont {N.}~\bibnamefont {Coppola}}, \bibinfo {author} {\bibfnamefont {F.}~\bibnamefont {Filsinger}}, \bibinfo {author} {\bibfnamefont {H.}~\bibnamefont {Gorke}}, \bibinfo {author} {\bibfnamefont {T.}~\bibnamefont
  {Gorkhover}}, \bibinfo {author} {\bibfnamefont {H.}~\bibnamefont {Graafsma}}, \bibinfo {author} {\bibfnamefont {L.}~\bibnamefont {Gumprecht}}, \bibinfo {author} {\bibfnamefont {A.}~\bibnamefont {Hartmann}}, \bibinfo {author} {\bibfnamefont {G.}~\bibnamefont {Hauser}}, \bibinfo {author} {\bibfnamefont {S.}~\bibnamefont {Herrmann}}, \bibinfo {author} {\bibfnamefont {H.}~\bibnamefont {Hirsemann}}, \bibinfo {author} {\bibfnamefont {P.}~\bibnamefont {Holl}}, \bibinfo {author} {\bibfnamefont {A.}~\bibnamefont {Hoemke}}, \bibinfo {author} {\bibfnamefont {L.}~\bibnamefont {Journel}}, \bibinfo {author} {\bibfnamefont {C.}~\bibnamefont {Kaiser}}, \bibinfo {author} {\bibfnamefont {N.}~\bibnamefont {Kimmel}}, \bibinfo {author} {\bibfnamefont {F.}~\bibnamefont {Krasniqi}}, \bibinfo {author} {\bibfnamefont {K.-U.}\ \bibnamefont {Kuehnel}}, \bibinfo {author} {\bibfnamefont {M.}~\bibnamefont {Matysek}}, \bibinfo {author} {\bibfnamefont {M.}~\bibnamefont {Messerschmidt}}, \bibinfo {author} {\bibfnamefont {D.}~\bibnamefont
  {Miesner}}, \bibinfo {author} {\bibfnamefont {T.}~\bibnamefont {Moeller}}, \bibinfo {author} {\bibfnamefont {R.}~\bibnamefont {Moshammer}}, \bibinfo {author} {\bibfnamefont {K.}~\bibnamefont {Nagaya}}, \bibinfo {author} {\bibfnamefont {B.}~\bibnamefont {Nilsson}}, \bibinfo {author} {\bibfnamefont {G.}~\bibnamefont {Potdevin}}, \bibinfo {author} {\bibfnamefont {D.}~\bibnamefont {Pietschner}}, \bibinfo {author} {\bibfnamefont {C.}~\bibnamefont {Reich}}, \bibinfo {author} {\bibfnamefont {D.}~\bibnamefont {Rupp}}, \bibinfo {author} {\bibfnamefont {G.}~\bibnamefont {Schaller}}, \bibinfo {author} {\bibfnamefont {I.}~\bibnamefont {Schlichting}}, \bibinfo {author} {\bibfnamefont {C.}~\bibnamefont {Schmidt}}, \bibinfo {author} {\bibfnamefont {F.}~\bibnamefont {Schopper}}, \bibinfo {author} {\bibfnamefont {S.}~\bibnamefont {Schorb}}, \bibinfo {author} {\bibfnamefont {C.-D.}\ \bibnamefont {Schroeter}}, \bibinfo {author} {\bibfnamefont {J.}~\bibnamefont {Schulz}}, \bibinfo {author} {\bibfnamefont {M.}~\bibnamefont
  {Simon}}, \bibinfo {author} {\bibfnamefont {H.}~\bibnamefont {Soltau}}, \bibinfo {author} {\bibfnamefont {L.}~\bibnamefont {Strueder}}, \bibinfo {author} {\bibfnamefont {K.}~\bibnamefont {Ueda}}, \bibinfo {author} {\bibfnamefont {G.}~\bibnamefont {Weidenspointner}}, \bibinfo {author} {\bibfnamefont {R.}~\bibnamefont {Santra}}, \bibinfo {author} {\bibfnamefont {J.}~\bibnamefont {Ullrich}}, \bibinfo {author} {\bibfnamefont {A.}~\bibnamefont {Rudenko}},\ and\ \bibinfo {author} {\bibfnamefont {D.}~\bibnamefont {Rolles}},\ }\bibfield  {title} {\bibinfo {title} {Ultra-efficient ionization of heavy atoms by intense x-ray free-electron laser pulses},\ }\href {https://doi.org/10.1038/Nphoton.2012.261} {\bibfield  {journal} {\bibinfo  {journal} {Nat. Photon.}\ }\textbf {\bibinfo {volume} {6}},\ \bibinfo {pages} {858} (\bibinfo {year} {2012})}\BibitemShut {NoStop}%
\bibitem [{\citenamefont {Doumy}\ \emph {et~al.}(2011)\citenamefont {Doumy}, \citenamefont {Roedig}, \citenamefont {Son}, \citenamefont {Blaga}, \citenamefont {DiChiara}, \citenamefont {Santra}, \citenamefont {Berrah}, \citenamefont {Bostedt}, \citenamefont {Bozek}, \citenamefont {Bucksbaum}, \citenamefont {Cryan}, \citenamefont {Fang}, \citenamefont {Ghimire}, \citenamefont {Glownia}, \citenamefont {Hoener}, \citenamefont {Kanter}, \citenamefont {Kr\"assig}, \citenamefont {Kuebel}, \citenamefont {Messerschmidt}, \citenamefont {Paulus}, \citenamefont {Reis}, \citenamefont {Rohringer}, \citenamefont {Young}, \citenamefont {Agostini},\ and\ \citenamefont {DiMauro}}]{Doumy-PRL-2011}%
  \BibitemOpen
  \bibfield  {author} {\bibinfo {author} {\bibfnamefont {G.}~\bibnamefont {Doumy}}, \bibinfo {author} {\bibfnamefont {C.}~\bibnamefont {Roedig}}, \bibinfo {author} {\bibfnamefont {S.-K.}\ \bibnamefont {Son}}, \bibinfo {author} {\bibfnamefont {C.~I.}\ \bibnamefont {Blaga}}, \bibinfo {author} {\bibfnamefont {A.~D.}\ \bibnamefont {DiChiara}}, \bibinfo {author} {\bibfnamefont {R.}~\bibnamefont {Santra}}, \bibinfo {author} {\bibfnamefont {N.}~\bibnamefont {Berrah}}, \bibinfo {author} {\bibfnamefont {C.}~\bibnamefont {Bostedt}}, \bibinfo {author} {\bibfnamefont {J.~D.}\ \bibnamefont {Bozek}}, \bibinfo {author} {\bibfnamefont {P.~H.}\ \bibnamefont {Bucksbaum}}, \bibinfo {author} {\bibfnamefont {J.~P.}\ \bibnamefont {Cryan}}, \bibinfo {author} {\bibfnamefont {L.}~\bibnamefont {Fang}}, \bibinfo {author} {\bibfnamefont {S.}~\bibnamefont {Ghimire}}, \bibinfo {author} {\bibfnamefont {J.~M.}\ \bibnamefont {Glownia}}, \bibinfo {author} {\bibfnamefont {M.}~\bibnamefont {Hoener}}, \bibinfo {author} {\bibfnamefont {E.~P.}\
  \bibnamefont {Kanter}}, \bibinfo {author} {\bibfnamefont {B.}~\bibnamefont {Kr\"assig}}, \bibinfo {author} {\bibfnamefont {M.}~\bibnamefont {Kuebel}}, \bibinfo {author} {\bibfnamefont {M.}~\bibnamefont {Messerschmidt}}, \bibinfo {author} {\bibfnamefont {G.~G.}\ \bibnamefont {Paulus}}, \bibinfo {author} {\bibfnamefont {D.~A.}\ \bibnamefont {Reis}}, \bibinfo {author} {\bibfnamefont {N.}~\bibnamefont {Rohringer}}, \bibinfo {author} {\bibfnamefont {L.}~\bibnamefont {Young}}, \bibinfo {author} {\bibfnamefont {P.}~\bibnamefont {Agostini}},\ and\ \bibinfo {author} {\bibfnamefont {L.~F.}\ \bibnamefont {DiMauro}},\ }\bibfield  {title} {\bibinfo {title} {Nonlinear atomic response to intense ultrashort x rays},\ }\href {https://doi.org/10.1103/PhysRevLett.106.083002} {\bibfield  {journal} {\bibinfo  {journal} {Phys. Rev. Lett.}\ }\textbf {\bibinfo {volume} {106}},\ \bibinfo {pages} {083002} (\bibinfo {year} {2011})}\BibitemShut {NoStop}%
\bibitem [{\citenamefont {Hoener}\ \emph {et~al.}(2010)\citenamefont {Hoener}, \citenamefont {Fang}, \citenamefont {Kornilov}, \citenamefont {Gessner}, \citenamefont {Pratt}, \citenamefont {G\"uhr}, \citenamefont {Kanter}, \citenamefont {Blaga}, \citenamefont {Bostedt}, \citenamefont {Bozek}, \citenamefont {Bucksbaum}, \citenamefont {Buth}, \citenamefont {Chen}, \citenamefont {Coffee}, \citenamefont {Cryan}, \citenamefont {DiMauro}, \citenamefont {Glownia}, \citenamefont {Hosler}, \citenamefont {Kukk}, \citenamefont {Leone}, \citenamefont {McFarland}, \citenamefont {Messerschmidt}, \citenamefont {Murphy}, \citenamefont {Petrovic}, \citenamefont {Rolles},\ and\ \citenamefont {Berrah}}]{Hoener-PRL-2010}%
  \BibitemOpen
  \bibfield  {author} {\bibinfo {author} {\bibfnamefont {M.}~\bibnamefont {Hoener}}, \bibinfo {author} {\bibfnamefont {L.}~\bibnamefont {Fang}}, \bibinfo {author} {\bibfnamefont {O.}~\bibnamefont {Kornilov}}, \bibinfo {author} {\bibfnamefont {O.}~\bibnamefont {Gessner}}, \bibinfo {author} {\bibfnamefont {S.~T.}\ \bibnamefont {Pratt}}, \bibinfo {author} {\bibfnamefont {M.}~\bibnamefont {G\"uhr}}, \bibinfo {author} {\bibfnamefont {E.~P.}\ \bibnamefont {Kanter}}, \bibinfo {author} {\bibfnamefont {C.}~\bibnamefont {Blaga}}, \bibinfo {author} {\bibfnamefont {C.}~\bibnamefont {Bostedt}}, \bibinfo {author} {\bibfnamefont {J.~D.}\ \bibnamefont {Bozek}}, \bibinfo {author} {\bibfnamefont {P.~H.}\ \bibnamefont {Bucksbaum}}, \bibinfo {author} {\bibfnamefont {C.}~\bibnamefont {Buth}}, \bibinfo {author} {\bibfnamefont {M.}~\bibnamefont {Chen}}, \bibinfo {author} {\bibfnamefont {R.}~\bibnamefont {Coffee}}, \bibinfo {author} {\bibfnamefont {J.}~\bibnamefont {Cryan}}, \bibinfo {author} {\bibfnamefont {L.}~\bibnamefont
  {DiMauro}}, \bibinfo {author} {\bibfnamefont {M.}~\bibnamefont {Glownia}}, \bibinfo {author} {\bibfnamefont {E.}~\bibnamefont {Hosler}}, \bibinfo {author} {\bibfnamefont {E.}~\bibnamefont {Kukk}}, \bibinfo {author} {\bibfnamefont {S.~R.}\ \bibnamefont {Leone}}, \bibinfo {author} {\bibfnamefont {B.}~\bibnamefont {McFarland}}, \bibinfo {author} {\bibfnamefont {M.}~\bibnamefont {Messerschmidt}}, \bibinfo {author} {\bibfnamefont {B.}~\bibnamefont {Murphy}}, \bibinfo {author} {\bibfnamefont {V.}~\bibnamefont {Petrovic}}, \bibinfo {author} {\bibfnamefont {D.}~\bibnamefont {Rolles}},\ and\ \bibinfo {author} {\bibfnamefont {N.}~\bibnamefont {Berrah}},\ }\bibfield  {title} {\bibinfo {title} {Ultraintense x-ray induced ionization, dissociation, and frustrated absorption in molecular nitrogen},\ }\href {https://doi.org/10.1103/PhysRevLett.104.253002} {\bibfield  {journal} {\bibinfo  {journal} {Phys. Rev. Lett.}\ }\textbf {\bibinfo {volume} {104}},\ \bibinfo {pages} {253002} (\bibinfo {year} {2010})}\BibitemShut
  {NoStop}%
\bibitem [{\citenamefont {Schorb}\ \emph {et~al.}(2012)\citenamefont {Schorb}, \citenamefont {Rupp}, \citenamefont {Swiggers}, \citenamefont {Coffee}, \citenamefont {Messerschmidt}, \citenamefont {Williams}, \citenamefont {Bozek}, \citenamefont {Wada}, \citenamefont {Kornilov}, \citenamefont {M\"oller},\ and\ \citenamefont {Bostedt}}]{Schorb-PRL-2012}%
  \BibitemOpen
  \bibfield  {author} {\bibinfo {author} {\bibfnamefont {S.}~\bibnamefont {Schorb}}, \bibinfo {author} {\bibfnamefont {D.}~\bibnamefont {Rupp}}, \bibinfo {author} {\bibfnamefont {M.~L.}\ \bibnamefont {Swiggers}}, \bibinfo {author} {\bibfnamefont {R.~N.}\ \bibnamefont {Coffee}}, \bibinfo {author} {\bibfnamefont {M.}~\bibnamefont {Messerschmidt}}, \bibinfo {author} {\bibfnamefont {G.}~\bibnamefont {Williams}}, \bibinfo {author} {\bibfnamefont {J.~D.}\ \bibnamefont {Bozek}}, \bibinfo {author} {\bibfnamefont {S.-I.}\ \bibnamefont {Wada}}, \bibinfo {author} {\bibfnamefont {O.}~\bibnamefont {Kornilov}}, \bibinfo {author} {\bibfnamefont {T.}~\bibnamefont {M\"oller}},\ and\ \bibinfo {author} {\bibfnamefont {C.}~\bibnamefont {Bostedt}},\ }\bibfield  {title} {\bibinfo {title} {Size-dependent ultrafast ionization dynamics of nanoscale samples in intense femtosecond x-ray free-electron-laser pulses},\ }\href {https://doi.org/10.1103/PhysRevLett.108.233401} {\bibfield  {journal} {\bibinfo  {journal} {Phys. Rev. Lett.}\
  }\textbf {\bibinfo {volume} {108}},\ \bibinfo {pages} {233401} (\bibinfo {year} {2012})}\BibitemShut {NoStop}%
\bibitem [{\citenamefont {Bostedt}\ \emph {et~al.}(2012)\citenamefont {Bostedt}, \citenamefont {Eremina}, \citenamefont {Rupp}, \citenamefont {Adolph}, \citenamefont {Thomas}, \citenamefont {Hoener}, \citenamefont {de~Castro}, \citenamefont {Tiggesb\"aumker}, \citenamefont {Meiwes-Broer}, \citenamefont {Laarmann}, \citenamefont {Wabnitz}, \citenamefont {Pl\"onjes}, \citenamefont {Treusch}, \citenamefont {Schneider},\ and\ \citenamefont {M\"oller}}]{Bostedt2012}%
  \BibitemOpen
  \bibfield  {author} {\bibinfo {author} {\bibfnamefont {C.}~\bibnamefont {Bostedt}}, \bibinfo {author} {\bibfnamefont {E.}~\bibnamefont {Eremina}}, \bibinfo {author} {\bibfnamefont {D.}~\bibnamefont {Rupp}}, \bibinfo {author} {\bibfnamefont {M.}~\bibnamefont {Adolph}}, \bibinfo {author} {\bibfnamefont {H.}~\bibnamefont {Thomas}}, \bibinfo {author} {\bibfnamefont {M.}~\bibnamefont {Hoener}}, \bibinfo {author} {\bibfnamefont {A.~R.~B.}\ \bibnamefont {de~Castro}}, \bibinfo {author} {\bibfnamefont {J.}~\bibnamefont {Tiggesb\"aumker}}, \bibinfo {author} {\bibfnamefont {K.-H.}\ \bibnamefont {Meiwes-Broer}}, \bibinfo {author} {\bibfnamefont {T.}~\bibnamefont {Laarmann}}, \bibinfo {author} {\bibfnamefont {H.}~\bibnamefont {Wabnitz}}, \bibinfo {author} {\bibfnamefont {E.}~\bibnamefont {Pl\"onjes}}, \bibinfo {author} {\bibfnamefont {R.}~\bibnamefont {Treusch}}, \bibinfo {author} {\bibfnamefont {J.~R.}\ \bibnamefont {Schneider}},\ and\ \bibinfo {author} {\bibfnamefont {T.}~\bibnamefont {M\"oller}},\ }\bibfield  {title}
  {\bibinfo {title} {Ultrafast x-ray scattering of xenon nanoparticles: Imaging transient states of matter},\ }\href {https://doi.org/10.1103/PhysRevLett.108.093401} {\bibfield  {journal} {\bibinfo  {journal} {Phys. Rev. Lett.}\ }\textbf {\bibinfo {volume} {108}},\ \bibinfo {pages} {093401} (\bibinfo {year} {2012})}\BibitemShut {NoStop}%
\bibitem [{\citenamefont {Inoue}\ \emph {et~al.}(2023)\citenamefont {Inoue}, \citenamefont {Yamada}, \citenamefont {Kapcia}, \citenamefont {Stransky}, \citenamefont {Tkachenko}, \citenamefont {Jurek}, \citenamefont {Inoue}, \citenamefont {Osaka}, \citenamefont {Inubushi}, \citenamefont {Ito}, \citenamefont {Tanaka}, \citenamefont {Matsuyama}, \citenamefont {Yamauchi}, \citenamefont {Yabashi},\ and\ \citenamefont {Ziaja}}]{Femtosecond_reduction_formfactor2023}%
  \BibitemOpen
  \bibfield  {author} {\bibinfo {author} {\bibfnamefont {I.}~\bibnamefont {Inoue}}, \bibinfo {author} {\bibfnamefont {J.}~\bibnamefont {Yamada}}, \bibinfo {author} {\bibfnamefont {K.~J.}\ \bibnamefont {Kapcia}}, \bibinfo {author} {\bibfnamefont {M.}~\bibnamefont {Stransky}}, \bibinfo {author} {\bibfnamefont {V.}~\bibnamefont {Tkachenko}}, \bibinfo {author} {\bibfnamefont {Z.}~\bibnamefont {Jurek}}, \bibinfo {author} {\bibfnamefont {T.}~\bibnamefont {Inoue}}, \bibinfo {author} {\bibfnamefont {T.}~\bibnamefont {Osaka}}, \bibinfo {author} {\bibfnamefont {Y.}~\bibnamefont {Inubushi}}, \bibinfo {author} {\bibfnamefont {A.}~\bibnamefont {Ito}}, \bibinfo {author} {\bibfnamefont {Y.}~\bibnamefont {Tanaka}}, \bibinfo {author} {\bibfnamefont {S.}~\bibnamefont {Matsuyama}}, \bibinfo {author} {\bibfnamefont {K.}~\bibnamefont {Yamauchi}}, \bibinfo {author} {\bibfnamefont {M.}~\bibnamefont {Yabashi}},\ and\ \bibinfo {author} {\bibfnamefont {B.}~\bibnamefont {Ziaja}},\ }\bibfield  {title} {\bibinfo {title} {Femtosecond
  reduction of atomic scattering factors triggered by intense x-ray pulse},\ }\href {https://doi.org/10.1103/PhysRevLett.131.163201} {\bibfield  {journal} {\bibinfo  {journal} {Phys. Rev. Lett.}\ }\textbf {\bibinfo {volume} {131}},\ \bibinfo {pages} {163201} (\bibinfo {year} {2023})}\BibitemShut {NoStop}%
\bibitem [{\citenamefont {Yumoto}\ \emph {et~al.}(2022)\citenamefont {Yumoto}, \citenamefont {Koyama}, \citenamefont {Suzuki}, \citenamefont {Joti}, \citenamefont {Niida}, \citenamefont {Tono}, \citenamefont {Bessho}, \citenamefont {Yabashi}, \citenamefont {Nishino},\ and\ \citenamefont {Ohashi}}]{Yumoto2022}%
  \BibitemOpen
  \bibfield  {author} {\bibinfo {author} {\bibfnamefont {H.}~\bibnamefont {Yumoto}}, \bibinfo {author} {\bibfnamefont {T.}~\bibnamefont {Koyama}}, \bibinfo {author} {\bibfnamefont {A.}~\bibnamefont {Suzuki}}, \bibinfo {author} {\bibfnamefont {Y.}~\bibnamefont {Joti}}, \bibinfo {author} {\bibfnamefont {Y.}~\bibnamefont {Niida}}, \bibinfo {author} {\bibfnamefont {K.}~\bibnamefont {Tono}}, \bibinfo {author} {\bibfnamefont {Y.}~\bibnamefont {Bessho}}, \bibinfo {author} {\bibfnamefont {M.}~\bibnamefont {Yabashi}}, \bibinfo {author} {\bibfnamefont {Y.}~\bibnamefont {Nishino}},\ and\ \bibinfo {author} {\bibfnamefont {H.}~\bibnamefont {Ohashi}},\ }\bibfield  {title} {\bibinfo {title} {High-fluence and high-gain multilayer focusing optics to enhance spatial resolution in femtosecond x-ray laser imaging},\ }\href@noop {} {\bibfield  {journal} {\bibinfo  {journal} {Nature communications}\ }\textbf {\bibinfo {volume} {13}},\ \bibinfo {pages} {1} (\bibinfo {year} {2022})}\BibitemShut {NoStop}%
\bibitem [{\citenamefont {Ekeberg}\ \emph {et~al.}(2024)\citenamefont {Ekeberg}, \citenamefont {Assalauova}, \citenamefont {Bielecki}, \citenamefont {Boll}, \citenamefont {Daurer}, \citenamefont {Eichacker}, \citenamefont {Franken}, \citenamefont {Galli}, \citenamefont {Gelisio}, \citenamefont {Gumprecht}, \citenamefont {Gunn}, \citenamefont {Hajdu}, \citenamefont {Hartmann}, \citenamefont {Hasse}, \citenamefont {Ignatenko}, \citenamefont {Koliyadu}, \citenamefont {Kulyk}, \citenamefont {Kurta}, \citenamefont {Kuster}, \citenamefont {Lugmayr}, \citenamefont {L{\"u}bke}, \citenamefont {Mancuso}, \citenamefont {Mazza}, \citenamefont {Nettelblad}, \citenamefont {Ovcharenko}, \citenamefont {Rivas}, \citenamefont {Rose}, \citenamefont {Samanta}, \citenamefont {Schmidt}, \citenamefont {Sobolev}, \citenamefont {Timneanu}, \citenamefont {Usenko}, \citenamefont {Westphal}, \citenamefont {Wollweber}, \citenamefont {Worbs}, \citenamefont {Xavier}, \citenamefont {Yousef}, \citenamefont {Ayyer}, \citenamefont {Chapman},
  \citenamefont {Sellberg}, \citenamefont {Seuring}, \citenamefont {Vartanyants}, \citenamefont {K{\"u}pper}, \citenamefont {Meyer},\ and\ \citenamefont {Maia}}]{Ekeberg2024}%
  \BibitemOpen
  \bibfield  {author} {\bibinfo {author} {\bibfnamefont {T.}~\bibnamefont {Ekeberg}}, \bibinfo {author} {\bibfnamefont {D.}~\bibnamefont {Assalauova}}, \bibinfo {author} {\bibfnamefont {J.}~\bibnamefont {Bielecki}}, \bibinfo {author} {\bibfnamefont {R.}~\bibnamefont {Boll}}, \bibinfo {author} {\bibfnamefont {B.~J.}\ \bibnamefont {Daurer}}, \bibinfo {author} {\bibfnamefont {L.~A.}\ \bibnamefont {Eichacker}}, \bibinfo {author} {\bibfnamefont {L.~E.}\ \bibnamefont {Franken}}, \bibinfo {author} {\bibfnamefont {D.~E.}\ \bibnamefont {Galli}}, \bibinfo {author} {\bibfnamefont {L.}~\bibnamefont {Gelisio}}, \bibinfo {author} {\bibfnamefont {L.}~\bibnamefont {Gumprecht}}, \bibinfo {author} {\bibfnamefont {L.~H.}\ \bibnamefont {Gunn}}, \bibinfo {author} {\bibfnamefont {J.}~\bibnamefont {Hajdu}}, \bibinfo {author} {\bibfnamefont {R.}~\bibnamefont {Hartmann}}, \bibinfo {author} {\bibfnamefont {D.}~\bibnamefont {Hasse}}, \bibinfo {author} {\bibfnamefont {A.}~\bibnamefont {Ignatenko}}, \bibinfo {author} {\bibfnamefont
  {J.}~\bibnamefont {Koliyadu}}, \bibinfo {author} {\bibfnamefont {O.}~\bibnamefont {Kulyk}}, \bibinfo {author} {\bibfnamefont {R.}~\bibnamefont {Kurta}}, \bibinfo {author} {\bibfnamefont {M.}~\bibnamefont {Kuster}}, \bibinfo {author} {\bibfnamefont {W.}~\bibnamefont {Lugmayr}}, \bibinfo {author} {\bibfnamefont {J.}~\bibnamefont {L{\"u}bke}}, \bibinfo {author} {\bibfnamefont {A.~P.}\ \bibnamefont {Mancuso}}, \bibinfo {author} {\bibfnamefont {T.}~\bibnamefont {Mazza}}, \bibinfo {author} {\bibfnamefont {C.}~\bibnamefont {Nettelblad}}, \bibinfo {author} {\bibfnamefont {Y.}~\bibnamefont {Ovcharenko}}, \bibinfo {author} {\bibfnamefont {D.~E.}\ \bibnamefont {Rivas}}, \bibinfo {author} {\bibfnamefont {M.}~\bibnamefont {Rose}}, \bibinfo {author} {\bibfnamefont {A.~K.}\ \bibnamefont {Samanta}}, \bibinfo {author} {\bibfnamefont {P.}~\bibnamefont {Schmidt}}, \bibinfo {author} {\bibfnamefont {E.}~\bibnamefont {Sobolev}}, \bibinfo {author} {\bibfnamefont {N.}~\bibnamefont {Timneanu}}, \bibinfo {author} {\bibfnamefont
  {S.}~\bibnamefont {Usenko}}, \bibinfo {author} {\bibfnamefont {D.}~\bibnamefont {Westphal}}, \bibinfo {author} {\bibfnamefont {T.}~\bibnamefont {Wollweber}}, \bibinfo {author} {\bibfnamefont {L.}~\bibnamefont {Worbs}}, \bibinfo {author} {\bibfnamefont {P.~L.}\ \bibnamefont {Xavier}}, \bibinfo {author} {\bibfnamefont {H.}~\bibnamefont {Yousef}}, \bibinfo {author} {\bibfnamefont {K.}~\bibnamefont {Ayyer}}, \bibinfo {author} {\bibfnamefont {H.~N.}\ \bibnamefont {Chapman}}, \bibinfo {author} {\bibfnamefont {J.~A.}\ \bibnamefont {Sellberg}}, \bibinfo {author} {\bibfnamefont {C.}~\bibnamefont {Seuring}}, \bibinfo {author} {\bibfnamefont {I.~A.}\ \bibnamefont {Vartanyants}}, \bibinfo {author} {\bibfnamefont {J.}~\bibnamefont {K{\"u}pper}}, \bibinfo {author} {\bibfnamefont {M.}~\bibnamefont {Meyer}},\ and\ \bibinfo {author} {\bibfnamefont {F.~R. N.~C.}\ \bibnamefont {Maia}},\ }\bibfield  {title} {\bibinfo {title} {Observation of a single protein by ultrafast x-ray diffraction},\ }\href
  {https://doi.org/10.1038/s41377-023-01352-7} {\bibfield  {journal} {\bibinfo  {journal} {Light: Science {\&} Applications}\ }\textbf {\bibinfo {volume} {13}},\ \bibinfo {pages} {15} (\bibinfo {year} {2024})}\BibitemShut {NoStop}%
\bibitem [{\citenamefont {Rafie-Zinedine}\ \emph {et~al.}(2024)\citenamefont {Rafie-Zinedine}, \citenamefont {Varma~Yenupuri}, \citenamefont {Worbs}, \citenamefont {Maia}, \citenamefont {Heymann}, \citenamefont {Schulz},\ and\ \citenamefont {Bielecki}}]{Rafie-Zinedine-JSR-2024}%
  \BibitemOpen
  \bibfield  {author} {\bibinfo {author} {\bibfnamefont {S.}~\bibnamefont {Rafie-Zinedine}}, \bibinfo {author} {\bibfnamefont {T.}~\bibnamefont {Varma~Yenupuri}}, \bibinfo {author} {\bibfnamefont {L.}~\bibnamefont {Worbs}}, \bibinfo {author} {\bibfnamefont {F.~R. N.~C.}\ \bibnamefont {Maia}}, \bibinfo {author} {\bibfnamefont {M.}~\bibnamefont {Heymann}}, \bibinfo {author} {\bibfnamefont {J.}~\bibnamefont {Schulz}},\ and\ \bibinfo {author} {\bibfnamefont {J.}~\bibnamefont {Bielecki}},\ }\bibfield  {title} {\bibinfo {title} {{Enhancing electrospray ionization efficiency for~particle transmission through an aerodynamic lens stack}},\ }\href {https://doi.org/10.1107/S1600577524000158} {\bibfield  {journal} {\bibinfo  {journal} {Journal of Synchrotron Radiation}\ }\textbf {\bibinfo {volume} {31}},\ \bibinfo {pages} {222} (\bibinfo {year} {2024})}\BibitemShut {NoStop}%
\bibitem [{\citenamefont {Zimmermann}\ \emph {et~al.}(2023)\citenamefont {Zimmermann}, \citenamefont {Beguet}, \citenamefont {Guthruf}, \citenamefont {Langbehn},\ and\ \citenamefont {Rupp}}]{Zimmermann2023}%
  \BibitemOpen
  \bibfield  {author} {\bibinfo {author} {\bibfnamefont {J.}~\bibnamefont {Zimmermann}}, \bibinfo {author} {\bibfnamefont {F.}~\bibnamefont {Beguet}}, \bibinfo {author} {\bibfnamefont {D.}~\bibnamefont {Guthruf}}, \bibinfo {author} {\bibfnamefont {B.}~\bibnamefont {Langbehn}},\ and\ \bibinfo {author} {\bibfnamefont {D.}~\bibnamefont {Rupp}},\ }\bibfield  {title} {\bibinfo {title} {Finding the semantic similarity in single-particle diffraction images using self-supervised contrastive projection learning},\ }\href {https://doi.org/10.1038/s41524-023-00966-0} {\bibfield  {journal} {\bibinfo  {journal} {npj Computational Materials}\ }\textbf {\bibinfo {volume} {9}},\ \bibinfo {pages} {24} (\bibinfo {year} {2023})}\BibitemShut {NoStop}%
\bibitem [{\citenamefont {Ayyer}\ \emph {et~al.}(2021)\citenamefont {Ayyer}, \citenamefont {Xavier}, \citenamefont {Bielecki}, \citenamefont {Shen}, \citenamefont {Daurer}, \citenamefont {Samanta}, \citenamefont {Awel}, \citenamefont {Bean}, \citenamefont {Barty}, \citenamefont {Bergemann}, \citenamefont {Ekeberg}, \citenamefont {Estillore}, \citenamefont {Fangohr}, \citenamefont {Giewekemeyer}, \citenamefont {Hunter}, \citenamefont {Karnevskiy}, \citenamefont {Kirian}, \citenamefont {Kirkwood}, \citenamefont {Kim}, \citenamefont {Koliyadu}, \citenamefont {Lange}, \citenamefont {Letrun}, \citenamefont {L\"{u}bke}, \citenamefont {Michelat}, \citenamefont {Morgan}, \citenamefont {Roth}, \citenamefont {Sato}, \citenamefont {Sikorski}, \citenamefont {Schulz}, \citenamefont {Spence}, \citenamefont {Vagovic}, \citenamefont {Wollweber}, \citenamefont {Worbs}, \citenamefont {Yefanov}, \citenamefont {Zhuang}, \citenamefont {Maia}, \citenamefont {Horke}, \citenamefont {K\"{u}pper}, \citenamefont {Loh}, \citenamefont
  {Mancuso},\ and\ \citenamefont {Chapman}}]{Ayyer:21}%
  \BibitemOpen
  \bibfield  {author} {\bibinfo {author} {\bibfnamefont {K.}~\bibnamefont {Ayyer}}, \bibinfo {author} {\bibfnamefont {P.~L.}\ \bibnamefont {Xavier}}, \bibinfo {author} {\bibfnamefont {J.}~\bibnamefont {Bielecki}}, \bibinfo {author} {\bibfnamefont {Z.}~\bibnamefont {Shen}}, \bibinfo {author} {\bibfnamefont {B.~J.}\ \bibnamefont {Daurer}}, \bibinfo {author} {\bibfnamefont {A.~K.}\ \bibnamefont {Samanta}}, \bibinfo {author} {\bibfnamefont {S.}~\bibnamefont {Awel}}, \bibinfo {author} {\bibfnamefont {R.}~\bibnamefont {Bean}}, \bibinfo {author} {\bibfnamefont {A.}~\bibnamefont {Barty}}, \bibinfo {author} {\bibfnamefont {M.}~\bibnamefont {Bergemann}}, \bibinfo {author} {\bibfnamefont {T.}~\bibnamefont {Ekeberg}}, \bibinfo {author} {\bibfnamefont {A.~D.}\ \bibnamefont {Estillore}}, \bibinfo {author} {\bibfnamefont {H.}~\bibnamefont {Fangohr}}, \bibinfo {author} {\bibfnamefont {K.}~\bibnamefont {Giewekemeyer}}, \bibinfo {author} {\bibfnamefont {M.~S.}\ \bibnamefont {Hunter}}, \bibinfo {author} {\bibfnamefont
  {M.}~\bibnamefont {Karnevskiy}}, \bibinfo {author} {\bibfnamefont {R.~A.}\ \bibnamefont {Kirian}}, \bibinfo {author} {\bibfnamefont {H.}~\bibnamefont {Kirkwood}}, \bibinfo {author} {\bibfnamefont {Y.}~\bibnamefont {Kim}}, \bibinfo {author} {\bibfnamefont {J.}~\bibnamefont {Koliyadu}}, \bibinfo {author} {\bibfnamefont {H.}~\bibnamefont {Lange}}, \bibinfo {author} {\bibfnamefont {R.}~\bibnamefont {Letrun}}, \bibinfo {author} {\bibfnamefont {J.}~\bibnamefont {L\"{u}bke}}, \bibinfo {author} {\bibfnamefont {T.}~\bibnamefont {Michelat}}, \bibinfo {author} {\bibfnamefont {A.~J.}\ \bibnamefont {Morgan}}, \bibinfo {author} {\bibfnamefont {N.}~\bibnamefont {Roth}}, \bibinfo {author} {\bibfnamefont {T.}~\bibnamefont {Sato}}, \bibinfo {author} {\bibfnamefont {M.}~\bibnamefont {Sikorski}}, \bibinfo {author} {\bibfnamefont {F.}~\bibnamefont {Schulz}}, \bibinfo {author} {\bibfnamefont {J.~C.~H.}\ \bibnamefont {Spence}}, \bibinfo {author} {\bibfnamefont {P.}~\bibnamefont {Vagovic}}, \bibinfo {author} {\bibfnamefont
  {T.}~\bibnamefont {Wollweber}}, \bibinfo {author} {\bibfnamefont {L.}~\bibnamefont {Worbs}}, \bibinfo {author} {\bibfnamefont {O.}~\bibnamefont {Yefanov}}, \bibinfo {author} {\bibfnamefont {Y.}~\bibnamefont {Zhuang}}, \bibinfo {author} {\bibfnamefont {F.~R. N.~C.}\ \bibnamefont {Maia}}, \bibinfo {author} {\bibfnamefont {D.~A.}\ \bibnamefont {Horke}}, \bibinfo {author} {\bibfnamefont {J.}~\bibnamefont {K\"{u}pper}}, \bibinfo {author} {\bibfnamefont {N.~D.}\ \bibnamefont {Loh}}, \bibinfo {author} {\bibfnamefont {A.~P.}\ \bibnamefont {Mancuso}},\ and\ \bibinfo {author} {\bibfnamefont {H.~N.}\ \bibnamefont {Chapman}},\ }\bibfield  {title} {\bibinfo {title} {3d diffractive imaging of nanoparticle ensembles using an x-ray laser},\ }\href {https://doi.org/10.1364/OPTICA.410851} {\bibfield  {journal} {\bibinfo  {journal} {Optica}\ }\textbf {\bibinfo {volume} {8}},\ \bibinfo {pages} {15} (\bibinfo {year} {2021})}\BibitemShut {NoStop}%
\bibitem [{\citenamefont {Marinelli}\ \emph {et~al.}(2017)\citenamefont {Marinelli}, \citenamefont {MacArthur}, \citenamefont {Emma}, \citenamefont {Guetg}, \citenamefont {Field}, \citenamefont {Kharakh}, \citenamefont {Lutman}, \citenamefont {Ding},\ and\ \citenamefont {Huang}}]{Marinelli-APL-2017}%
  \BibitemOpen
  \bibfield  {author} {\bibinfo {author} {\bibfnamefont {A.}~\bibnamefont {Marinelli}}, \bibinfo {author} {\bibfnamefont {J.}~\bibnamefont {MacArthur}}, \bibinfo {author} {\bibfnamefont {P.}~\bibnamefont {Emma}}, \bibinfo {author} {\bibfnamefont {M.}~\bibnamefont {Guetg}}, \bibinfo {author} {\bibfnamefont {C.}~\bibnamefont {Field}}, \bibinfo {author} {\bibfnamefont {D.}~\bibnamefont {Kharakh}}, \bibinfo {author} {\bibfnamefont {A.~A.}\ \bibnamefont {Lutman}}, \bibinfo {author} {\bibfnamefont {Y.}~\bibnamefont {Ding}},\ and\ \bibinfo {author} {\bibfnamefont {Z.}~\bibnamefont {Huang}},\ }\bibfield  {title} {\bibinfo {title} {{Experimental demonstration of a single-spike hard-X-ray free-electron laser starting from noise}},\ }\href {https://doi.org/10.1063/1.4990716} {\bibfield  {journal} {\bibinfo  {journal} {Applied Physics Letters}\ }\textbf {\bibinfo {volume} {111}},\ \bibinfo {pages} {151101} (\bibinfo {year} {2017})},\ \Eprint
  {https://arxiv.org/abs/https://pubs.aip.org/aip/apl/article-pdf/doi/10.1063/1.4990716/13155677/151101\_1\_online.pdf} {https://pubs.aip.org/aip/apl/article-pdf/doi/10.1063/1.4990716/13155677/151101\_1\_online.pdf} \BibitemShut {NoStop}%
\bibitem [{\citenamefont {Huang}\ \emph {et~al.}(2017)\citenamefont {Huang}, \citenamefont {Ding}, \citenamefont {Feng}, \citenamefont {Hemsing}, \citenamefont {Huang}, \citenamefont {Krzywinski}, \citenamefont {Lutman}, \citenamefont {Marinelli}, \citenamefont {Maxwell},\ and\ \citenamefont {Zhu}}]{Huang-PRL-2017}%
  \BibitemOpen
  \bibfield  {author} {\bibinfo {author} {\bibfnamefont {S.}~\bibnamefont {Huang}}, \bibinfo {author} {\bibfnamefont {Y.}~\bibnamefont {Ding}}, \bibinfo {author} {\bibfnamefont {Y.}~\bibnamefont {Feng}}, \bibinfo {author} {\bibfnamefont {E.}~\bibnamefont {Hemsing}}, \bibinfo {author} {\bibfnamefont {Z.}~\bibnamefont {Huang}}, \bibinfo {author} {\bibfnamefont {J.}~\bibnamefont {Krzywinski}}, \bibinfo {author} {\bibfnamefont {A.~A.}\ \bibnamefont {Lutman}}, \bibinfo {author} {\bibfnamefont {A.}~\bibnamefont {Marinelli}}, \bibinfo {author} {\bibfnamefont {T.~J.}\ \bibnamefont {Maxwell}},\ and\ \bibinfo {author} {\bibfnamefont {D.}~\bibnamefont {Zhu}},\ }\bibfield  {title} {\bibinfo {title} {Generating single-spike hard x-ray pulses with nonlinear bunch compression in free-electron lasers},\ }\href {https://doi.org/10.1103/PhysRevLett.119.154801} {\bibfield  {journal} {\bibinfo  {journal} {Phys. Rev. Lett.}\ }\textbf {\bibinfo {volume} {119}},\ \bibinfo {pages} {154801} (\bibinfo {year} {2017})}\BibitemShut
  {NoStop}%
\bibitem [{\citenamefont {Duris}\ \emph {et~al.}(2020)\citenamefont {Duris}, \citenamefont {Li}, \citenamefont {Driver}, \citenamefont {Champenois}, \citenamefont {MacArthur}, \citenamefont {Lutman}, \citenamefont {Zhang}, \citenamefont {Rosenberger}, \citenamefont {Aldrich}, \citenamefont {Coffee} \emph {et~al.}}]{attoXLEAP2020}%
  \BibitemOpen
  \bibfield  {author} {\bibinfo {author} {\bibfnamefont {J.}~\bibnamefont {Duris}}, \bibinfo {author} {\bibfnamefont {S.}~\bibnamefont {Li}}, \bibinfo {author} {\bibfnamefont {T.}~\bibnamefont {Driver}}, \bibinfo {author} {\bibfnamefont {E.~G.}\ \bibnamefont {Champenois}}, \bibinfo {author} {\bibfnamefont {J.~P.}\ \bibnamefont {MacArthur}}, \bibinfo {author} {\bibfnamefont {A.~A.}\ \bibnamefont {Lutman}}, \bibinfo {author} {\bibfnamefont {Z.}~\bibnamefont {Zhang}}, \bibinfo {author} {\bibfnamefont {P.}~\bibnamefont {Rosenberger}}, \bibinfo {author} {\bibfnamefont {J.~W.}\ \bibnamefont {Aldrich}}, \bibinfo {author} {\bibfnamefont {R.}~\bibnamefont {Coffee}}, \emph {et~al.},\ }\bibfield  {title} {\bibinfo {title} {Tunable isolated attosecond x-ray pulses with gigawatt peak power from a free-electron laser},\ }\href@noop {} {\bibfield  {journal} {\bibinfo  {journal} {Nature Photonics}\ }\textbf {\bibinfo {volume} {14}},\ \bibinfo {pages} {30} (\bibinfo {year} {2020})}\BibitemShut {NoStop}%
\bibitem [{\citenamefont {Malyzhenkov}\ \emph {et~al.}(2020)\citenamefont {Malyzhenkov}, \citenamefont {Arbelo}, \citenamefont {Craievich}, \citenamefont {Dijkstal}, \citenamefont {Ferrari}, \citenamefont {Reiche}, \citenamefont {Schietinger}, \citenamefont {Jurani\ifmmode~\acute{c}\else \'{c}\fi{}},\ and\ \citenamefont {Prat}}]{Malyzhenkov-2020-PRR}%
  \BibitemOpen
  \bibfield  {author} {\bibinfo {author} {\bibfnamefont {A.}~\bibnamefont {Malyzhenkov}}, \bibinfo {author} {\bibfnamefont {Y.~P.}\ \bibnamefont {Arbelo}}, \bibinfo {author} {\bibfnamefont {P.}~\bibnamefont {Craievich}}, \bibinfo {author} {\bibfnamefont {P.}~\bibnamefont {Dijkstal}}, \bibinfo {author} {\bibfnamefont {E.}~\bibnamefont {Ferrari}}, \bibinfo {author} {\bibfnamefont {S.}~\bibnamefont {Reiche}}, \bibinfo {author} {\bibfnamefont {T.}~\bibnamefont {Schietinger}}, \bibinfo {author} {\bibfnamefont {P.}~\bibnamefont {Jurani\ifmmode~\acute{c}\else \'{c}\fi{}}},\ and\ \bibinfo {author} {\bibfnamefont {E.}~\bibnamefont {Prat}},\ }\bibfield  {title} {\bibinfo {title} {Single- and two-color attosecond hard x-ray free-electron laser pulses with nonlinear compression},\ }\href {https://doi.org/10.1103/PhysRevResearch.2.042018} {\bibfield  {journal} {\bibinfo  {journal} {Phys. Rev. Res.}\ }\textbf {\bibinfo {volume} {2}},\ \bibinfo {pages} {042018} (\bibinfo {year} {2020})}\BibitemShut {NoStop}%
\bibitem [{\citenamefont {Trebushinin}\ \emph {et~al.}(2023)\citenamefont {Trebushinin}, \citenamefont {Geloni}, \citenamefont {Serkez}, \citenamefont {Mercurio}, \citenamefont {Gerasimova}, \citenamefont {Maltezopoulos}, \citenamefont {Guetg},\ and\ \citenamefont {Schneidmiller}}]{Trebushinin-Photonics-2023}%
  \BibitemOpen
  \bibfield  {author} {\bibinfo {author} {\bibfnamefont {A.}~\bibnamefont {Trebushinin}}, \bibinfo {author} {\bibfnamefont {G.}~\bibnamefont {Geloni}}, \bibinfo {author} {\bibfnamefont {S.}~\bibnamefont {Serkez}}, \bibinfo {author} {\bibfnamefont {G.}~\bibnamefont {Mercurio}}, \bibinfo {author} {\bibfnamefont {N.}~\bibnamefont {Gerasimova}}, \bibinfo {author} {\bibfnamefont {T.}~\bibnamefont {Maltezopoulos}}, \bibinfo {author} {\bibfnamefont {M.}~\bibnamefont {Guetg}},\ and\ \bibinfo {author} {\bibfnamefont {E.}~\bibnamefont {Schneidmiller}},\ }\bibfield  {title} {\bibinfo {title} {Experimental demonstration of attoseconds-at-harmonics at the sase3 undulator of the european xfel},\ }\bibfield  {journal} {\bibinfo  {journal} {Photonics}\ }\textbf {\bibinfo {volume} {10}},\ \href {https://doi.org/10.3390/photonics10020131} {10.3390/photonics10020131} (\bibinfo {year} {2023})\BibitemShut {NoStop}%
\bibitem [{\citenamefont {Nandi}\ \emph {et~al.}(2022)\citenamefont {Nandi}, \citenamefont {Olofsson}, \citenamefont {Bertolino}, \citenamefont {Carlstr{\"o}m}, \citenamefont {Zapata}, \citenamefont {Busto}, \citenamefont {Callegari}, \citenamefont {Di~Fraia}, \citenamefont {Eng-Johnsson}, \citenamefont {Feifel} \emph {et~al.}}]{nandi2022_observationRabidynamics}%
  \BibitemOpen
  \bibfield  {author} {\bibinfo {author} {\bibfnamefont {S.}~\bibnamefont {Nandi}}, \bibinfo {author} {\bibfnamefont {E.}~\bibnamefont {Olofsson}}, \bibinfo {author} {\bibfnamefont {M.}~\bibnamefont {Bertolino}}, \bibinfo {author} {\bibfnamefont {S.}~\bibnamefont {Carlstr{\"o}m}}, \bibinfo {author} {\bibfnamefont {F.}~\bibnamefont {Zapata}}, \bibinfo {author} {\bibfnamefont {D.}~\bibnamefont {Busto}}, \bibinfo {author} {\bibfnamefont {C.}~\bibnamefont {Callegari}}, \bibinfo {author} {\bibfnamefont {M.}~\bibnamefont {Di~Fraia}}, \bibinfo {author} {\bibfnamefont {P.}~\bibnamefont {Eng-Johnsson}}, \bibinfo {author} {\bibfnamefont {R.}~\bibnamefont {Feifel}}, \emph {et~al.},\ }\bibfield  {title} {\bibinfo {title} {Observation of rabi dynamics with a short-wavelength free-electron laser},\ }\href@noop {} {\bibfield  {journal} {\bibinfo  {journal} {Nature}\ }\textbf {\bibinfo {volume} {608}},\ \bibinfo {pages} {488} (\bibinfo {year} {2022})}\BibitemShut {NoStop}%
\bibitem [{\citenamefont {Zhang}\ \emph {et~al.}(2024{\natexlab{a}})\citenamefont {Zhang}, \citenamefont {Liao}, \citenamefont {Chen}, \citenamefont {Yu}, \citenamefont {Li}, \citenamefont {He}, \citenamefont {Liu}, \citenamefont {Lu},\ and\ \citenamefont {Zhou}}]{Zhang-2024-PRA}%
  \BibitemOpen
  \bibfield  {author} {\bibinfo {author} {\bibfnamefont {X.}~\bibnamefont {Zhang}}, \bibinfo {author} {\bibfnamefont {Y.}~\bibnamefont {Liao}}, \bibinfo {author} {\bibfnamefont {Y.}~\bibnamefont {Chen}}, \bibinfo {author} {\bibfnamefont {M.}~\bibnamefont {Yu}}, \bibinfo {author} {\bibfnamefont {S.}~\bibnamefont {Li}}, \bibinfo {author} {\bibfnamefont {M.}~\bibnamefont {He}}, \bibinfo {author} {\bibfnamefont {K.}~\bibnamefont {Liu}}, \bibinfo {author} {\bibfnamefont {P.}~\bibnamefont {Lu}},\ and\ \bibinfo {author} {\bibfnamefont {Y.}~\bibnamefont {Zhou}},\ }\bibfield  {title} {\bibinfo {title} {Monitoring the ultrafast buildup of rabi oscillations in the time domain},\ }\href {https://doi.org/10.1103/PhysRevA.109.063102} {\bibfield  {journal} {\bibinfo  {journal} {Phys. Rev. A}\ }\textbf {\bibinfo {volume} {109}},\ \bibinfo {pages} {063102} (\bibinfo {year} {2024}{\natexlab{a}})}\BibitemShut {NoStop}%
\bibitem [{\citenamefont {Zhang}\ \emph {et~al.}(2024{\natexlab{b}})\citenamefont {Zhang}, \citenamefont {Liao}, \citenamefont {Chen}, \citenamefont {Yu}, \citenamefont {Liu}, \citenamefont {Lu},\ and\ \citenamefont {Zhou}}]{Zhang-2024-PRA2}%
  \BibitemOpen
  \bibfield  {author} {\bibinfo {author} {\bibfnamefont {X.}~\bibnamefont {Zhang}}, \bibinfo {author} {\bibfnamefont {Y.}~\bibnamefont {Liao}}, \bibinfo {author} {\bibfnamefont {Y.}~\bibnamefont {Chen}}, \bibinfo {author} {\bibfnamefont {M.}~\bibnamefont {Yu}}, \bibinfo {author} {\bibfnamefont {K.}~\bibnamefont {Liu}}, \bibinfo {author} {\bibfnamefont {P.}~\bibnamefont {Lu}},\ and\ \bibinfo {author} {\bibfnamefont {Y.}~\bibnamefont {Zhou}},\ }\bibfield  {title} {\bibinfo {title} {Characterization of the two-photon transition phase in the buildup process of rabi oscillations},\ }\href {https://doi.org/10.1103/PhysRevA.109.033101} {\bibfield  {journal} {\bibinfo  {journal} {Phys. Rev. A}\ }\textbf {\bibinfo {volume} {109}},\ \bibinfo {pages} {033101} (\bibinfo {year} {2024}{\natexlab{b}})}\BibitemShut {NoStop}%
\bibitem [{\citenamefont {Kanter}\ \emph {et~al.}(2011)\citenamefont {Kanter}, \citenamefont {Kr\"assig}, \citenamefont {Li}, \citenamefont {March}, \citenamefont {Ho}, \citenamefont {Rohringer}, \citenamefont {Santra}, \citenamefont {Southworth}, \citenamefont {DiMauro}, \citenamefont {Doumy}, \citenamefont {Roedig}, \citenamefont {Berrah}, \citenamefont {Fang}, \citenamefont {Hoener}, \citenamefont {Bucksbaum}, \citenamefont {Ghimire}, \citenamefont {Reis}, \citenamefont {Bozek}, \citenamefont {Bostedt}, \citenamefont {Messerschmidt},\ and\ \citenamefont {Young}}]{Hiddenresonance_kanter}%
  \BibitemOpen
  \bibfield  {author} {\bibinfo {author} {\bibfnamefont {E.~P.}\ \bibnamefont {Kanter}}, \bibinfo {author} {\bibfnamefont {B.}~\bibnamefont {Kr\"assig}}, \bibinfo {author} {\bibfnamefont {Y.}~\bibnamefont {Li}}, \bibinfo {author} {\bibfnamefont {A.~M.}\ \bibnamefont {March}}, \bibinfo {author} {\bibfnamefont {P.}~\bibnamefont {Ho}}, \bibinfo {author} {\bibfnamefont {N.}~\bibnamefont {Rohringer}}, \bibinfo {author} {\bibfnamefont {R.}~\bibnamefont {Santra}}, \bibinfo {author} {\bibfnamefont {S.~H.}\ \bibnamefont {Southworth}}, \bibinfo {author} {\bibfnamefont {L.~F.}\ \bibnamefont {DiMauro}}, \bibinfo {author} {\bibfnamefont {G.}~\bibnamefont {Doumy}}, \bibinfo {author} {\bibfnamefont {C.~A.}\ \bibnamefont {Roedig}}, \bibinfo {author} {\bibfnamefont {N.}~\bibnamefont {Berrah}}, \bibinfo {author} {\bibfnamefont {L.}~\bibnamefont {Fang}}, \bibinfo {author} {\bibfnamefont {M.}~\bibnamefont {Hoener}}, \bibinfo {author} {\bibfnamefont {P.~H.}\ \bibnamefont {Bucksbaum}}, \bibinfo {author} {\bibfnamefont
  {S.}~\bibnamefont {Ghimire}}, \bibinfo {author} {\bibfnamefont {D.~A.}\ \bibnamefont {Reis}}, \bibinfo {author} {\bibfnamefont {J.~D.}\ \bibnamefont {Bozek}}, \bibinfo {author} {\bibfnamefont {C.}~\bibnamefont {Bostedt}}, \bibinfo {author} {\bibfnamefont {M.}~\bibnamefont {Messerschmidt}},\ and\ \bibinfo {author} {\bibfnamefont {L.}~\bibnamefont {Young}},\ }\bibfield  {title} {\bibinfo {title} {Unveiling and driving hidden resonances with high-fluence, high-intensity x-ray pulses},\ }\href {https://doi.org/10.1103/PhysRevLett.107.233001} {\bibfield  {journal} {\bibinfo  {journal} {Phys. Rev. Lett.}\ }\textbf {\bibinfo {volume} {107}},\ \bibinfo {pages} {233001} (\bibinfo {year} {2011})}\BibitemShut {NoStop}%
\bibitem [{\citenamefont {Kuschel}\ \emph {et~al.}(2022)\citenamefont {Kuschel}, \citenamefont {Ho}, \citenamefont {Haddad}, \citenamefont {Zimmermann}, \citenamefont {Flueckiger}, \citenamefont {Ware}, \citenamefont {Duris}, \citenamefont {MacArthur}, \citenamefont {Lutman}, \citenamefont {Lin} \emph {et~al.}}]{Tais_2022_Xenon_preprint}%
  \BibitemOpen
  \bibfield  {author} {\bibinfo {author} {\bibfnamefont {S.}~\bibnamefont {Kuschel}}, \bibinfo {author} {\bibfnamefont {P.~J.}\ \bibnamefont {Ho}}, \bibinfo {author} {\bibfnamefont {A.~A.}\ \bibnamefont {Haddad}}, \bibinfo {author} {\bibfnamefont {F.}~\bibnamefont {Zimmermann}}, \bibinfo {author} {\bibfnamefont {L.}~\bibnamefont {Flueckiger}}, \bibinfo {author} {\bibfnamefont {M.~R.}\ \bibnamefont {Ware}}, \bibinfo {author} {\bibfnamefont {J.}~\bibnamefont {Duris}}, \bibinfo {author} {\bibfnamefont {J.~P.}\ \bibnamefont {MacArthur}}, \bibinfo {author} {\bibfnamefont {A.}~\bibnamefont {Lutman}}, \bibinfo {author} {\bibfnamefont {M.-F.}\ \bibnamefont {Lin}}, \emph {et~al.},\ }\bibfield  {title} {\bibinfo {title} {Enhanced ultrafast x-ray diffraction by transient resonances},\ }\href@noop {} {\bibfield  {journal} {\bibinfo  {journal} {arXiv preprint arXiv:2207.05472}\ } (\bibinfo {year} {2022})}\BibitemShut {NoStop}%
\bibitem [{\citenamefont {Ulmer}\ \emph {et~al.}(2023)\citenamefont {Ulmer}, \citenamefont {Kuschel}, \citenamefont {Langbehn}, \citenamefont {Hecht}, \citenamefont {Dold}, \citenamefont {R{\"o}nnebeck}, \citenamefont {Driver}, \citenamefont {Duris}, \citenamefont {Kamalov}, \citenamefont {Li} \emph {et~al.}}]{Tais_Ne_apstalk}%
  \BibitemOpen
  \bibfield  {author} {\bibinfo {author} {\bibfnamefont {A.}~\bibnamefont {Ulmer}}, \bibinfo {author} {\bibfnamefont {S.}~\bibnamefont {Kuschel}}, \bibinfo {author} {\bibfnamefont {B.}~\bibnamefont {Langbehn}}, \bibinfo {author} {\bibfnamefont {L.}~\bibnamefont {Hecht}}, \bibinfo {author} {\bibfnamefont {S.}~\bibnamefont {Dold}}, \bibinfo {author} {\bibfnamefont {L.}~\bibnamefont {R{\"o}nnebeck}}, \bibinfo {author} {\bibfnamefont {T.}~\bibnamefont {Driver}}, \bibinfo {author} {\bibfnamefont {J.}~\bibnamefont {Duris}}, \bibinfo {author} {\bibfnamefont {A.}~\bibnamefont {Kamalov}}, \bibinfo {author} {\bibfnamefont {X.}~\bibnamefont {Li}}, \emph {et~al.},\ }\bibfield  {title} {\bibinfo {title} {Exploring damage reduction and scattering cross section enhancement in attosecond x-ray imaging of neon near the k-edge},\ }\href@noop {} {\bibfield  {journal} {\bibinfo  {journal} {Bulletin of the American Physical Society}\ } (\bibinfo {year} {2023})}\BibitemShut {NoStop}%
\bibitem [{\citenamefont {Brown}\ and\ \citenamefont {Kibble}(1964)}]{KB}%
  \BibitemOpen
  \bibfield  {author} {\bibinfo {author} {\bibfnamefont {L.~S.}\ \bibnamefont {Brown}}\ and\ \bibinfo {author} {\bibfnamefont {T.~W.~B.}\ \bibnamefont {Kibble}},\ }\bibfield  {title} {\bibinfo {title} {Interaction of intense laser beams with electrons},\ }\href {https://doi.org/10.1103/PhysRev.133.A705} {\bibfield  {journal} {\bibinfo  {journal} {Phys. Rev.}\ }\textbf {\bibinfo {volume} {133}},\ \bibinfo {pages} {A705} (\bibinfo {year} {1964})}\BibitemShut {NoStop}%
\bibitem [{\citenamefont {Venkatesh}\ and\ \citenamefont {Robicheaux}(2020)}]{NLCPRA_1}%
  \BibitemOpen
  \bibfield  {author} {\bibinfo {author} {\bibfnamefont {A.}~\bibnamefont {Venkatesh}}\ and\ \bibinfo {author} {\bibfnamefont {F.}~\bibnamefont {Robicheaux}},\ }\bibfield  {title} {\bibinfo {title} {Simulation of nonlinear {C}ompton scattering from bound electrons},\ }\href {https://doi.org/10.1103/PhysRevA.101.013409} {\bibfield  {journal} {\bibinfo  {journal} {Phys. Rev. A}\ }\textbf {\bibinfo {volume} {101}},\ \bibinfo {pages} {013409} (\bibinfo {year} {2020})}\BibitemShut {NoStop}%
\bibitem [{\citenamefont {Venkatesh}\ and\ \citenamefont {Robicheaux}(2022)}]{NLCPRA_4}%
  \BibitemOpen
  \bibfield  {author} {\bibinfo {author} {\bibfnamefont {A.}~\bibnamefont {Venkatesh}}\ and\ \bibinfo {author} {\bibfnamefont {F.}~\bibnamefont {Robicheaux}},\ }\bibfield  {title} {\bibinfo {title} {Simulations for x-ray imaging of wave-packet dynamics},\ }\href {https://doi.org/10.1103/PhysRevA.106.033125} {\bibfield  {journal} {\bibinfo  {journal} {Phys. Rev. A}\ }\textbf {\bibinfo {volume} {106}},\ \bibinfo {pages} {033125} (\bibinfo {year} {2022})}\BibitemShut {NoStop}%
\bibitem [{Com(2024)}]{Companion_article}%
  \BibitemOpen
  \href@noop {} {\bibinfo {title} {Companion article}} (\bibinfo {year} {2024})\BibitemShut {NoStop}%
\bibitem [{\citenamefont {Vewinger}\ \emph {et~al.}(2003)\citenamefont {Vewinger}, \citenamefont {Heinz}, \citenamefont {Garcia~Fernandez}, \citenamefont {Vitanov},\ and\ \citenamefont {Bergmann}}]{Bergman_Neon_superposition_PRL2003}%
  \BibitemOpen
  \bibfield  {author} {\bibinfo {author} {\bibfnamefont {F.}~\bibnamefont {Vewinger}}, \bibinfo {author} {\bibfnamefont {M.}~\bibnamefont {Heinz}}, \bibinfo {author} {\bibfnamefont {R.}~\bibnamefont {Garcia~Fernandez}}, \bibinfo {author} {\bibfnamefont {N.~V.}\ \bibnamefont {Vitanov}},\ and\ \bibinfo {author} {\bibfnamefont {K.}~\bibnamefont {Bergmann}},\ }\bibfield  {title} {\bibinfo {title} {Creation and measurement of a coherent superposition of quantum states},\ }\href {https://doi.org/10.1103/PhysRevLett.91.213001} {\bibfield  {journal} {\bibinfo  {journal} {Phys. Rev. Lett.}\ }\textbf {\bibinfo {volume} {91}},\ \bibinfo {pages} {213001} (\bibinfo {year} {2003})}\BibitemShut {NoStop}%
\bibitem [{\citenamefont {Vewinger}\ \emph {et~al.}(2007{\natexlab{a}})\citenamefont {Vewinger}, \citenamefont {Heinz}, \citenamefont {Shore},\ and\ \citenamefont {Bergmann}}]{Bergman_Neon_superposition_PRA2007_theory}%
  \BibitemOpen
  \bibfield  {author} {\bibinfo {author} {\bibfnamefont {F.}~\bibnamefont {Vewinger}}, \bibinfo {author} {\bibfnamefont {M.}~\bibnamefont {Heinz}}, \bibinfo {author} {\bibfnamefont {B.~W.}\ \bibnamefont {Shore}},\ and\ \bibinfo {author} {\bibfnamefont {K.}~\bibnamefont {Bergmann}},\ }\bibfield  {title} {\bibinfo {title} {Amplitude and phase control of a coherent superposition of degenerate states. i. theory},\ }\href {https://doi.org/10.1103/PhysRevA.75.043406} {\bibfield  {journal} {\bibinfo  {journal} {Phys. Rev. A}\ }\textbf {\bibinfo {volume} {75}},\ \bibinfo {pages} {043406} (\bibinfo {year} {2007}{\natexlab{a}})}\BibitemShut {NoStop}%
\bibitem [{\citenamefont {Vewinger}\ \emph {et~al.}(2007{\natexlab{b}})\citenamefont {Vewinger}, \citenamefont {Heinz}, \citenamefont {Schneider}, \citenamefont {Barthel},\ and\ \citenamefont {Bergmann}}]{Bergman_Neon_superposition_PRA2007_exp}%
  \BibitemOpen
  \bibfield  {author} {\bibinfo {author} {\bibfnamefont {F.}~\bibnamefont {Vewinger}}, \bibinfo {author} {\bibfnamefont {M.}~\bibnamefont {Heinz}}, \bibinfo {author} {\bibfnamefont {U.}~\bibnamefont {Schneider}}, \bibinfo {author} {\bibfnamefont {C.}~\bibnamefont {Barthel}},\ and\ \bibinfo {author} {\bibfnamefont {K.}~\bibnamefont {Bergmann}},\ }\bibfield  {title} {\bibinfo {title} {Amplitude and phase control of a coherent superposition of degenerate states. ii. experiment},\ }\href {https://doi.org/10.1103/PhysRevA.75.043407} {\bibfield  {journal} {\bibinfo  {journal} {Phys. Rev. A}\ }\textbf {\bibinfo {volume} {75}},\ \bibinfo {pages} {043407} (\bibinfo {year} {2007}{\natexlab{b}})}\BibitemShut {NoStop}%
\bibitem [{\citenamefont {Eberly}(1998)}]{Pulsearea_defn_Eberly}%
  \BibitemOpen
  \bibfield  {author} {\bibinfo {author} {\bibfnamefont {J.}~\bibnamefont {Eberly}},\ }\bibfield  {title} {\bibinfo {title} {Area theorem rederived},\ }\href {https://doi.org/10.1364/OE.2.000173} {\bibfield  {journal} {\bibinfo  {journal} {Opt. Express}\ }\textbf {\bibinfo {volume} {2}},\ \bibinfo {pages} {173} (\bibinfo {year} {1998})}\BibitemShut {NoStop}%
\bibitem [{\citenamefont {Cavaletto}\ \emph {et~al.}(2012)\citenamefont {Cavaletto}, \citenamefont {Buth}, \citenamefont {Harman}, \citenamefont {Kanter}, \citenamefont {Southworth}, \citenamefont {Young},\ and\ \citenamefont {Keitel}}]{Cavaletto_ResFluor_PRA}%
  \BibitemOpen
  \bibfield  {author} {\bibinfo {author} {\bibfnamefont {S.~M.}\ \bibnamefont {Cavaletto}}, \bibinfo {author} {\bibfnamefont {C.}~\bibnamefont {Buth}}, \bibinfo {author} {\bibfnamefont {Z.}~\bibnamefont {Harman}}, \bibinfo {author} {\bibfnamefont {E.~P.}\ \bibnamefont {Kanter}}, \bibinfo {author} {\bibfnamefont {S.~H.}\ \bibnamefont {Southworth}}, \bibinfo {author} {\bibfnamefont {L.}~\bibnamefont {Young}},\ and\ \bibinfo {author} {\bibfnamefont {C.~H.}\ \bibnamefont {Keitel}},\ }\bibfield  {title} {\bibinfo {title} {Resonance fluorescence in ultrafast and intense x-ray free-electron-laser pulses},\ }\href {https://doi.org/10.1103/PhysRevA.86.033402} {\bibfield  {journal} {\bibinfo  {journal} {Phys. Rev. A}\ }\textbf {\bibinfo {volume} {86}},\ \bibinfo {pages} {033402} (\bibinfo {year} {2012})}\BibitemShut {NoStop}%
\bibitem [{\citenamefont {Fischer}\ \emph {et~al.}(2017)\citenamefont {Fischer}, \citenamefont {Hanschke}, \citenamefont {Kremser}, \citenamefont {Finley}, \citenamefont {M{\"u}ller},\ and\ \citenamefont {Vu{\v{c}}kovi{\'c}}}]{pulseareatheorem_2018}%
  \BibitemOpen
  \bibfield  {author} {\bibinfo {author} {\bibfnamefont {K.~A.}\ \bibnamefont {Fischer}}, \bibinfo {author} {\bibfnamefont {L.}~\bibnamefont {Hanschke}}, \bibinfo {author} {\bibfnamefont {M.}~\bibnamefont {Kremser}}, \bibinfo {author} {\bibfnamefont {J.~J.}\ \bibnamefont {Finley}}, \bibinfo {author} {\bibfnamefont {K.}~\bibnamefont {M{\"u}ller}},\ and\ \bibinfo {author} {\bibfnamefont {J.}~\bibnamefont {Vu{\v{c}}kovi{\'c}}},\ }\bibfield  {title} {\bibinfo {title} {Pulsed rabi oscillations in quantum two-level systems: beyond the area theorem},\ }\href@noop {} {\bibfield  {journal} {\bibinfo  {journal} {Quantum Science and Technology}\ }\textbf {\bibinfo {volume} {3}},\ \bibinfo {pages} {014006} (\bibinfo {year} {2017})}\BibitemShut {NoStop}%
\bibitem [{\citenamefont {Sakurai}(1967)}]{Sakurai_adv}%
  \BibitemOpen
  \bibfield  {author} {\bibinfo {author} {\bibfnamefont {J.}~\bibnamefont {Sakurai}},\ }\href@noop {} {\emph {\bibinfo {title} {Advanced Quantum Mechanics}}}\ (\bibinfo  {publisher} {Addison-Wesley Publishing Company},\ \bibinfo {year} {1967})\BibitemShut {NoStop}%
\bibitem [{\citenamefont {Bhowmick}\ \emph {et~al.}(2023)\citenamefont {Bhowmick}, \citenamefont {Simon}, \citenamefont {Bogacz}, \citenamefont {Hussein}, \citenamefont {Zhang}, \citenamefont {Makita}, \citenamefont {Ibrahim}, \citenamefont {Chatterjee}, \citenamefont {Doyle}, \citenamefont {Cheah}, \citenamefont {Chernev}, \citenamefont {Fuller}, \citenamefont {Fransson}, \citenamefont {Alonso-Mori}, \citenamefont {Brewster}, \citenamefont {Sauter}, \citenamefont {Bergmann}, \citenamefont {Dobbek}, \citenamefont {Zouni}, \citenamefont {Messinger}, \citenamefont {Kern}, \citenamefont {Yachandra},\ and\ \citenamefont {Yano}}]{Bhowmick-IUCRJ-2023}%
  \BibitemOpen
  \bibfield  {author} {\bibinfo {author} {\bibfnamefont {A.}~\bibnamefont {Bhowmick}}, \bibinfo {author} {\bibfnamefont {P.~S.}\ \bibnamefont {Simon}}, \bibinfo {author} {\bibfnamefont {I.}~\bibnamefont {Bogacz}}, \bibinfo {author} {\bibfnamefont {R.}~\bibnamefont {Hussein}}, \bibinfo {author} {\bibfnamefont {M.}~\bibnamefont {Zhang}}, \bibinfo {author} {\bibfnamefont {H.}~\bibnamefont {Makita}}, \bibinfo {author} {\bibfnamefont {M.}~\bibnamefont {Ibrahim}}, \bibinfo {author} {\bibfnamefont {R.}~\bibnamefont {Chatterjee}}, \bibinfo {author} {\bibfnamefont {M.~D.}\ \bibnamefont {Doyle}}, \bibinfo {author} {\bibfnamefont {M.~H.}\ \bibnamefont {Cheah}}, \bibinfo {author} {\bibfnamefont {P.}~\bibnamefont {Chernev}}, \bibinfo {author} {\bibfnamefont {F.~D.}\ \bibnamefont {Fuller}}, \bibinfo {author} {\bibfnamefont {T.}~\bibnamefont {Fransson}}, \bibinfo {author} {\bibfnamefont {R.}~\bibnamefont {Alonso-Mori}}, \bibinfo {author} {\bibfnamefont {A.~S.}\ \bibnamefont {Brewster}}, \bibinfo {author} {\bibfnamefont
  {N.~K.}\ \bibnamefont {Sauter}}, \bibinfo {author} {\bibfnamefont {U.}~\bibnamefont {Bergmann}}, \bibinfo {author} {\bibfnamefont {H.}~\bibnamefont {Dobbek}}, \bibinfo {author} {\bibfnamefont {A.}~\bibnamefont {Zouni}}, \bibinfo {author} {\bibfnamefont {J.}~\bibnamefont {Messinger}}, \bibinfo {author} {\bibfnamefont {J.}~\bibnamefont {Kern}}, \bibinfo {author} {\bibfnamefont {V.~K.}\ \bibnamefont {Yachandra}},\ and\ \bibinfo {author} {\bibfnamefont {J.}~\bibnamefont {Yano}},\ }\bibfield  {title} {\bibinfo {title} {{Going around the Kok cycle of the water oxidation reaction with femtosecond X-ray crystallography}},\ }\href {https://doi.org/10.1107/S2052252523008928} {\bibfield  {journal} {\bibinfo  {journal} {IUCrJ}\ }\textbf {\bibinfo {volume} {10}},\ \bibinfo {pages} {642} (\bibinfo {year} {2023})}\BibitemShut {NoStop}%
\end{thebibliography}%

\end{document}